\documentclass[authoryear,preprint,12pt]{elsarticle}

\usepackage[english]{babel}
\usepackage[utf8x]{inputenc}
\usepackage[T1]{fontenc}

\usepackage[a4paper,top=3cm,bottom=2cm,left=3cm,right=3cm,marginparwidth=1.75cm]{geometry}

\usepackage{amsmath}
\usepackage{censor}
\usepackage{graphicx}
\usepackage{subfigure}

\usepackage{python}
\usepackage[colorlinks=true, allcolors=blue]{hyperref}

\newcommand{\new}[1]{{#1}}

\begin{document}

\title{Detecting social (in)stability in primates from their temporal co-presence network}

\author[cpt]{Valeria Gelardi}
\author[lpc]{Jo\"el Fagot} 
\author[cpt,isi]{Alain Barrat}
\author[lpc,corresp]{Nicolas Claidi\`ere}

\address[cpt]{Aix Marseille Univ, Universit\'e de Toulon, CNRS, CPT, Marseille, France}
\address[lpc]{Aix Marseille Univ, CNRS, LPC, FED3C, Marseille, France}
\address[isi]{Data Science Laboratory, ISI Foundation, Turin, Italy}
\address[corresp]{Corresponding author: Nicolas Claidière, Laboratoire de Psychologie Cognitive, Aix Marseille Université, 3 Place Victor Hugo, 13331 Marseille, France. Email: nicolas.claidiere@normalesup.org Phone: 0033 695 358 417}

\begin{abstract}  
 {The stability of social relationships is  {important} to animals living in groups, and social network analysis provides a powerful tool to help characterize and understand their (in)stability and the consequences at the group level. However, the use of dynamic social networks is still limited 
  {in this context}
 because it requires long-term 
social data and new analytical tools.}
 {Here, we study the dynamic evolution of a group of 29 Guinea baboons ({\it Papio papio}) using a dataset of automatically collected cognitive tests comprising more than 16M records collected over 3 years.} We first built a monthly aggregated temporal network describing the baboon’s co-presence in the cognitive testing booths. We then used a null model, considering the heterogeneity in the baboons' activity, to define both positive (association) and negative (avoidance) 
monthly networks.
 {We tested social balance theory by combining
these positive and negative social networks. The results showed that the networks were structurally balanced and that newly created edges 
also tended to preserve social balance.} We then investigated several network metrics to gain insights into the  {individual level and group level social networks long-term temporal evolution}. Interestingly, a measure of similarity between successive monthly networks was able to pinpoint periods of stability and instability and to show how some baboons' ego-networks remained stable while others changed radically.
 {Our study confirms the prediction of social balance theory but also shows that 
  {large fluctuations in the numbers of triads may limit its} applicability 
to study the dynamic evolution of animal social networks. In contrast, the use of the similarity measure proved to be very versatile and sensitive in detecting relationships' (in)stabilities at different levels. The changes  we identified can be linked, at least in some cases, to females changing primary male, as observed in the wild.}

\end{abstract}

\maketitle



\section*{INTRODUCTION}
 The stability of social relationships is  {important} to group living animals and is a significant aspect of social structure \citep{Hinde:1976}. In primates societies for instance, stable and long-lasting relationships can enhance individuals fitness \citep{silk2007social,Silk:2010,Alberts2019} through offspring survival \citep{Silk:2003,Silk:2009} and reproduction \citep{Schulke:2010}. 
  \new{There is also evidence that disruption of social stability can have negative consequences, in zebra finches for instance, see \citep{maldonado-chaparro:2018}.}  
\new{Despite the benefits of
stability in relationships, 
natural processes such as demography naturally
bring changes in 
 social relationships \citep{shizuka:2019}: 
 individuals thus adapt and} flexibly change
 their social strategies on different timescales depending on context, from a daily basis  \citep{Sick:2014}, to seasonal \citep{Henzi:2009} to several years \citep{Silk:2010}. The  relationship
 changes, which can be based for instance on factors such as genetic relatedness \citep{Beisner:2011} and personality \citep{McCowan:2011}, can in turn give rise to stability or instability at the group level.

Social network analysis is a powerful tool to better understand the dynamic changes in social relationships because it provides a conceptual framework {that uses dyadic interactions to infer social relationships and group level properties} \citep{Hinde:1976,Hobson:2013,Kurvers:2014,wey2008social,Croft:2008,Krause:2015,Whitehead:2008, WEBBER201977}. 
\new{To go further and analyse the dynamic changes in social relationships and their consequences at the group level,} a dynamic social network analysis (dynSNA) approach is needed. However, the use of \new{a dynamic network
approach} is still limited for non-human animals  \citep{Pinter-Wollman:2014}. 
In many cases progress in this direction has been hampered by the nature of the techniques used to gather social information  \citep{Farine:2015}, although recent technological developments are starting to provide automatically collected high-resolution data  \citep{Hughey:2018}. 
 
\new{The nature and evolution of each dyadic relationship 
could of course be studied by itself: it can for instance 
be positive, negative, appear or disappear, and
its change in case of a social instability can be
investigated. 
Studying each relation in isolation however is not 
sufficient, as the evolution of a dyadic relationship is most
often related to the other dyads of the group. 
To understand the evolution
of single dyads,
the whole network evolution needs thus to be considered,
as well as the structures formed by dyads at various
scales, such as triangles. 
Few tools are however yet available to study network 
structures 
and their evolution at various scales,  
 from individuals' dyadic relationships to 
 intermediate scales and to the entire social network.} 
  {In the present work, our goal is to contribute to bridging this gap by presenting analytical tools and a methodological framework to study the dynamics of animal
 temporal networks and in particular to detect periods of stability and instability.}

 {The simplest social structures beyond dyadic associations involve three individuals, i.e., triads, among which at most three links can exist.} Triadic closure {refers} to the fact that, if only two links are known to be present in a triad (e.g., between three individuals A B and C, only the links AB and BC are known to exist), the third link AC is predicted to exist as well or to appear in the future. This phenomenon helps to characterize and predict the development of ties within a network and the progression of its connectivity \citep{thurner2018virtual,Rapoport:1953,Granovetter:1973}. Evidence of triadic closure in primates has been reported by \cite{Borgeaud:2016} in three groups of vervet monkeys, finding that two individuals are more likely to be associated if they are both linked with a mutual third party associate. The notion of triadic closure is however limited to binary associations (a link is either present or not) and does not consider that links can be weighted or of different types (see however \cite{Brandenberger:2019}  for a recent extension to multi-edge social networks).

\begin{figure}[hb]
\centering
{\includegraphics[width=0.7\textwidth]{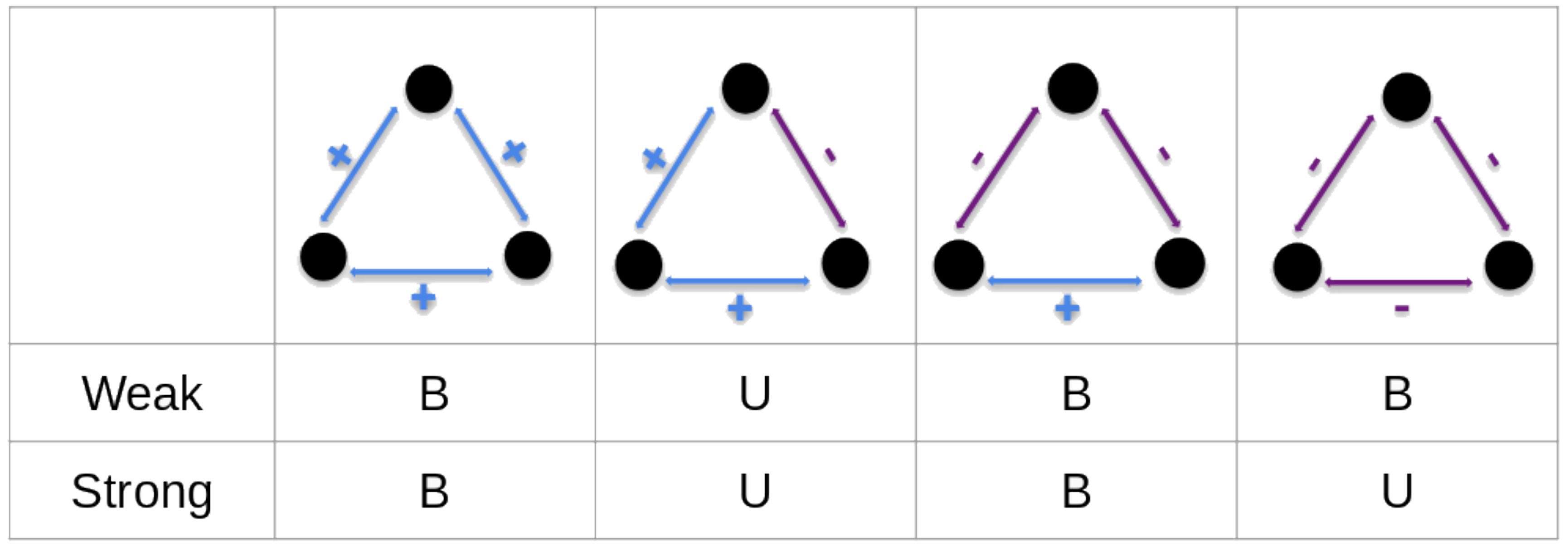}}
\caption{{\bf Principle of social balance theory.} 
\new{We show the four possible signed triads.
In the original "strong" formulation of the 
social balance theory, two triads are 
balanced (B) and two are unbalanced (U).
In the "weak" formulation put forward by \cite{davis1967clustering}, triangles with three negative edges are also considered balanced because 
such configurations can arise when more than two subgroups exist within the social network under consideration.
}}
\label{fig:socbalance_sketch}
\end{figure}

 {Triadic structures become richer}
when it is possible to assign signs (positive or negative) to the links of a social network, to denote on the one hand positive associations such as friendship or trust, and on the other hand antagonistic relationships such as dislike, distrust or aggression. In this case, social balance theory
\citep{heider1946attitudes, heider1958psychology,wasserman1994social,cartwright1956structural}
provides a theoretical framework to understand the dynamic (in)stability of signed social networks  {by studying the closed triangles},
which have each either 0, 1, 2 or 3 negative links (see Fig. \ref{fig:socbalance_sketch}).
In its strong formulation, social balance theory states that a triangle is balanced if 
all three links are positive (the three nodes are all "friends"), or if two links are negative and one is positive (two "friends" are "non-friends" with the third). The remaining two configurations (either three negative links or two positive links and one negative, see Fig. \ref{fig:socbalance_sketch}) are called unbalanced and are considered to be a symptom of tension and social stress. For instance, the configuration with two positive links and one negative is in contradiction with the common belief that "a friend of a friend is a friend".
\new{A network 
should thus} tend towards more balanced configurations \citep{Kulakowski:2005,Marvel:2011} as, 
in Heider's words, "If no balanced state exists, then forces towards this state will arise". 
\new{Studying social balance is therefore an interesting 
direction for investigating the dynamics of social
networks, as the 
numbers of balanced and unbalanced triads could be expected 
to be predictors of changes and instability.}
 
{Social balance theory has received a lot of theoretical attention 
\citep{Marvel:2011,Kulakowski:2005}
but relatively little empirical validation in human
social networks
\citep{Doreian:1996,Doreian:2009,Szell:2010,Leskovec:2010,thurner2018virtual}.
 {This is  most} probably due to the difficulty in defining signed social networks, i.e., in having information on both affiliative (positive) and antagonistic 
(negative) ties between individuals. 
The study of social balance theory in non-human animals is 
even more
difficult, as their social networks are often built using only information of positive nature (such as proximity)
 {and avoidance behaviour is difficult to observe and ascertain}. 
We know of only one study that explicitly tested social balance in a wild, non-human system.  In social rock hyrax ({\it Procavia capensis}), \cite{ilany2013structural}
considered two individuals who did not share a positive interaction within a year to be 'nonfriends' (negative link), and individuals who shared at least one positive interaction to be friends (positive link). By counting the number of balanced and unbalanced triangles, the authors confirmed the predictions of social balance theory in its strong version. They also showed that unbalanced triangles tended to change to become balanced, as predicted by the theory, but that some level of imbalance remained in the network, mostly due to new individuals entering the group and forming unbalanced structures.}

 {Social balance theory provides a very interesting framework as it formulates predictions both on the state of a social network and on the evolution of structures involving three individuals. However, it is limited to the smallest structures beyond dyads, namely triads. More flexible approaches to characterize dynamic changes across different levels of description of the social network (pairs, triads, subgroups of individuals, entire network) are thus needed and have started to be developed in the context of temporal networks \citep{holme2015modern,Fournet:2014,Darst:2016,Croft:2016,Farine:2017Dynamic}. Temporal networks representations of the evolution of social relationships are often composed of successive snapshots corresponding to the relationships observed or measured in successive time windows (e.g., successive weeks or months), and tools to detect changes between these time windows are typically based on measures of similarity between networks. Such measures can be used to detect the multiple time scales on which a network changes \citep{Darst:2016}; they can highlight abrupt changes of the network structure, or on the contrary periods during which the structure remains stable. They are also easily generalised to weighted networks \citep{Fournet:2014} and quantify the amount of structural and weights change of the network. }

 To develop data-driven investigations of dynamic social networks in animals, primates are particularly  {useful} 
 because many non-human primates have highly developed social relationships. For instance, they can use deceptive tactics \citep{Whiten:1988}, 
 be 
 prosocial \citep{Claidiere:2015},
 understand other's intentions \citep{Tomasello:2005}
 and can evaluate the potential helpfulness of others \citep{Anderson:2013}. Moreover, these social relationships are both structured and flexible. In our study species for example, Guinea baboons ({\it Papio papio}) exhibits complex social relationships and a multilevel social organization. The core unit of Guinea baboon's society is one primary male with 1−4 females and their offsprings; several core-units form parties, which form larger assemblies called gangs \citep{Patzelt:2014}. 

Males form strong bonds predominantly within parties; these bonds are not correlated with genetic relatedness and males are highly tolerant of each other \citep{Patzelt:2014}. 
Moreover, a recent study found that the social structure was flexible, with half of the females of a group having changed primary male at least once in 17 months \citep{Goffe:2016}, a behaviour also observed informally in our study group. Overall, a single group of individuals is thus in fact composed of multiple core units with a certain diversity of social structures, and the social network describing this social structure changes spontaneously. These 
characteristics make Guinea baboons ideal to study dynamic changes in social relationships.

 {In the present study, we leveraged a long-term dataset 
  {with high temporal resolution}
 collected automatically on a group of  {Guinea baboons}. We developed general analytical tools that can be used to study the structure and temporal evolution of non-human animal networks, with a particular emphasis on detecting the periods of stability and instability in the social network evolution (both at the individual and global level). We first put forward a systematic way to use the data to generate signed social networks among the individuals: to this aim, we defined both affiliative (positive) and antagonistic (negative) links between individuals, by comparing the data to a suitable null model \citep{Manly:1997,Bejder:1998,Farine:2017}.
 
This allowed us to investigate social balance theory in the social network of these non-human primates. We then considered various methods to detect changes in the network structure, either using tools linked to social balance theory, such as counts of the numbers of triangles and triangle creations, or through quantitative similarity measures between networks built in successive periods. In particular, the similarity metrics we considered allowed us to investigate the rearrangements of the network at various scales. It can indeed be measured at the scale of the whole network but also at more detailed levels such as subgroups or for each single individual. We highlight in particular how the local ego-network of some individuals could be completely altered while others were left unchanged, with only partial rearrangements of the overall social network. Notably, some of the periods of  interest and rearrangements of the ego-network of some individuals revealed by this analysis were confirmed by external observations.}

\section*{METHODS}

\subsection*{Participants}

Between January 2014 and May 2017, $29$ Guinea baboons (\emph{Papio papio}) belonging to a large social group {of the CNRS Primate Center in Rousset-sur-Arc (France)} participated in cognitive tests using Automatic Learning Devices for monkeys (ALDM). 
The size of the group varied from $19$ to $24$ individuals, because of several births and natural deaths during these years.
 {The monthly average size was} of $21.8$ individuals with $7.3$ $[7; 9]$ males and $14.4$ $[12; 17]$ females (mean [min; max]), with age ranging from $0$ to $21$ years old. The baboons were all marked by two biocompatible $1.2 \times 0.2$ cm RFID microchips injected into each forearm  {to individually identify each participant}.

\subsection*{Ethical note}

The baboons lived in an outdoor enclosure ($700 m^2$) connected to an indoor area that provided shelter when necessary. The outside enclosure was connected to ten testing booths freely accessible to the animals at any time where they could voluntarily perform ALDM
tests.  {This procedure reduces stress levels {\citep{fagot2014effects}}.} Water was provided ad libitum
within the enclosure, and they received their normal
ratio of food (fruits, vegetables, and monkey chow) every day
at 5 pm. The baboons were all born within the primate centre.
This research was carried out in accordance {with French standards and received approval from the national French ethics committee, the "Comit\'e d'Ethique CE-14 pour l'Exp\'erimentation Animale" (approval number APAFIS\#2717-2015111708173794)}. Procedures were also consistent with the guidelines of the Association for the Study of Animal Behaviour.

\subsection*{ALDM data}

\paragraph{ALDM database}

The dataset analysed here contains all recorded cognitive tests performed by the group of baboons {in a facility developed by J. Fagot \citep{Fagot2010,Fagot_Paleress_2009}}. In this facility baboons can freely access 10 workstations 
installed in two trailers (5 workstations in each trailer) connected to their enclosure. Each workstation is constituted of a test chamber with open rear side and transparent sidewalls. The front of the test chamber is fitted with a view port and two hand ports. By looking through the view port, participants can interact with an LCD touch screen installed at eye level and whenever a monkey introduces its forearm through one arm port, its RFID identity is recognized and triggers the presentation of the cognitive task on the touch screen. For each test, the date and time (with millisecond precision), the nature of the task, the name, age, sex and maternal family of the individual performing the test, and the identification number of the workstation in which the individual performed the test, is recorded. The system provides rewards in grains of dry wheat, delivered through a food dispenser, whenever the baboons succeed at the task. 

Crucially, in the ALDM system the monkeys cannot see each other's screen (observational learning is thus impossible),
  {but they have visual access to each other in neighbouring workstations through their transparent sides: they can thus approach booths together
 and see each other during the tests, meaning that co-presence may be interpreted as association.}
 {In fact,  {\cite{Claidiere2017}} recently showed that the timings of cognitive tests could be used to construct a social network of baboons highly similar to the one obtained from behavioural observations.}

The data covers the period from January 2014 to May 2017. The facility was however closed in the months of August 2014 and 2015, which are thus missing from the data. Furthermore, two individuals  {were born} in May 2017 so they do not appear in some of the analysis (e.g., in the measures of stability from one month to the next).

 {The data analysed here concern a total of 16 403 680 cognitive tests, representing an average of 13 186 records a day over 1 244 days, and  an average of 565 640 records per individual.}
 {The duration of a single bout of trials (i.e. a succession of trials  {of a given individual} separated by less than 5 seconds) was on average 60 seconds.}

\paragraph{Construction of the co-presence network: from raw data to monthly networks}

\begin{figure}[htb]
\centering
\includegraphics[width=0.7\textwidth]{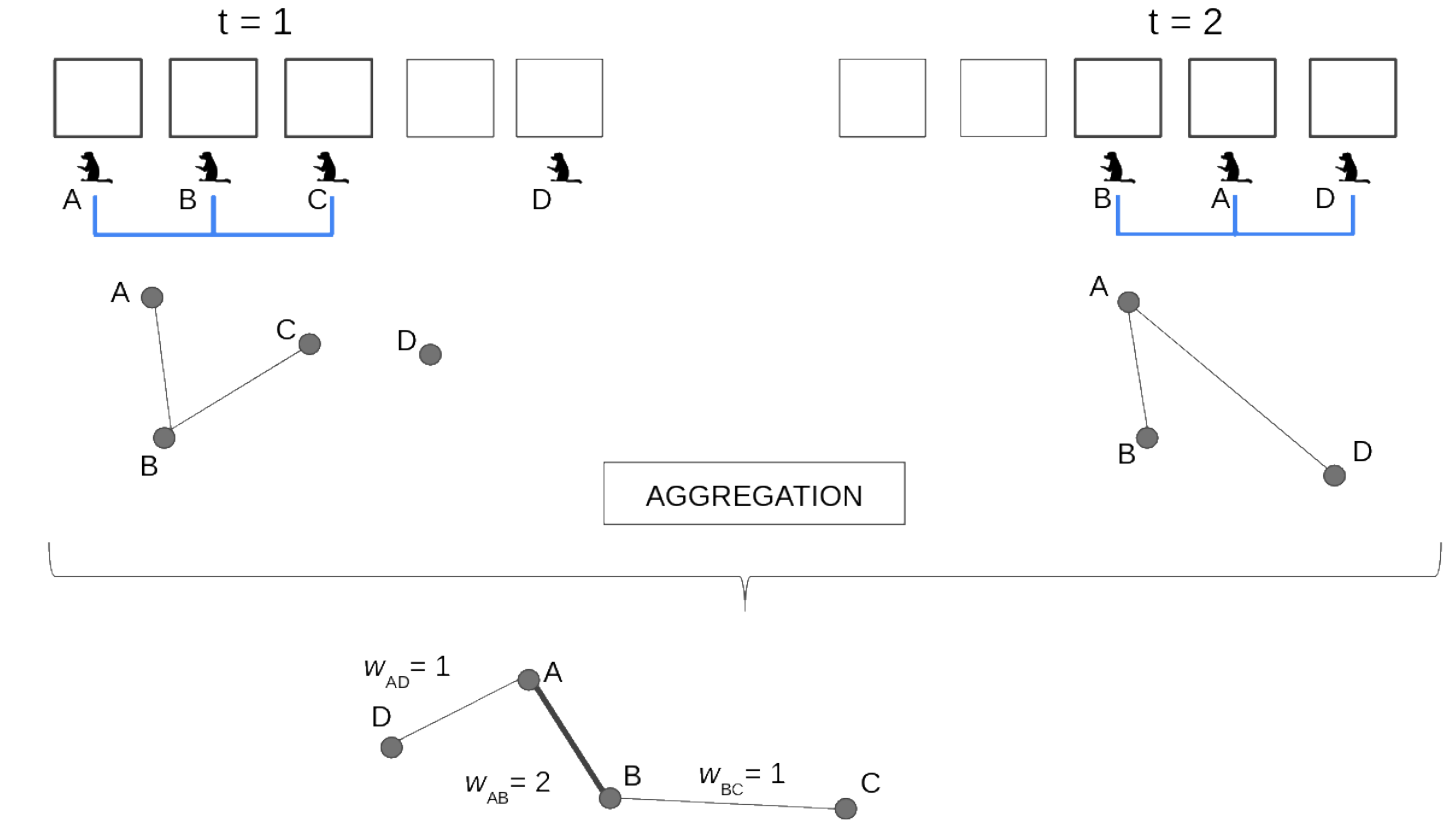} 
\caption{{\bf Sketch of the construction of the co-presence network}
Top: Schematic of the network construction from the ALDM dataset. In each time-window of $\Delta t = 5$ seconds, links were created between individuals (nodes) if they were recorded in adjacent workstations, as depicted in the middle drawings: links $AB$ and $BC$ in the first time-window, and $AB$ and $AD$ in the second. Afterwards, the instantaneous networks were aggregated to produce a weighted co-presence network (bottom sketch).} 
\label{fig:subfig_copres_net:construction}
\end{figure}

Using the ALDM dataset we built a temporal co-presence network based on the temporal and spatial proximity of the baboons (see Fig. \ref{fig:subfig_copres_net:construction}).
To this aim, we first
aggregated the raw ALDM data in successive temporal windows of $\Delta t = 5$ seconds: this interval length was short enough to consider that the individuals performing tests in the same time window were in co-presence, and long enough to have a sufficient number of co-presence events.
This interval length was also used by  {\cite{Claidiere2017}}, and we show in the
Appendix that our main results are robust with respect to changes in $\Delta t$.
  {Note that a single co-presence event lasting
for instance $1$ minute gave rise to $12$ successive time-windows in which the individuals were detected in co-presence}.

For each time window, we built a proximity network in which nodes represented baboons, and a link was drawn between two individuals if their presence had been recorded in two adjacent booths. We then aggregated these proximity networks on a monthly timescale, shown by  {\cite{Claidiere2017}} to correspond to an adequate aggregation timescale: in each monthly network, nodes represented baboons and 
a weighted link between two baboons represented that they had been in co-presence at least once; the weight of the link corresponded to the number of time windows in which co-presence of these individuals had been recorded
 {(i.e., the weight represented the total co-presence time, in units of 5 seconds)}.

 {\paragraph{Constructing a null model for the co-presence monthly network}
We considered the ALDM data of each month, which consists of the list of tests performed in that month,
with for each test the date, identity of the individual performing the test and identification of the workstation (see above).
In this list, we reshuffled the identities of the individuals among the tests, swapping all observations at once and keeping
the number of occurrences of each individual.
This procedure ensured that the frequency of participation of each individual and the spatial and temporal organisation of the trials 
were preserved  {for each month} 
in the null model \citep{Manly:1997,Bejder:1998,Farine:2017}. We then used the randomized data to create a weighted co-presence network as previously. We performed the 
randomization procedure $100$ times. 
Comparison of the co-presence networks with these randomized data allowed us to build monthly signed networks from the ALDM data as described below
in Results.
}

 {
\subsection*{Network metrics and evolution measures} A large variety of metrics have
been defined to describe the nodes of a network and the network as a whole. 
The simplest metric to characterize a node is given by its degree, i.e.,  the number of links to which it participates
or in other words its number of neighbors. Going beyond dyads, the clustering coefficient of a node quantifies the cohesiveness
of its neighborhood; it is defined for node $i$ as
\begin{equation}
    c_i = \frac{\Delta_i}{k_i(k_i-1)/2},
\end{equation}
where $k_i$ is the degree of node $i$, $\Delta_i$ is the number of closed triangles
$ijk$ to which $i$ participated (i.e., such that all links $ij$, $ik$ and $jk$ exist),  and the denominator gives the maximum possible
number of such triangles, ensuring that $c_i$ is 
bounded between $0$ and $1$. The average clustering coefficient of a network is then simply the average of $c_i$ over all the nodes. 
The transitivity of a network \citep{wasserman1994social} is also often used and gives a global quantification of its
cohesiveness through the ratio of the number of closed triangles divided by the number of triads.}

 {\paragraph{Balanced and unbalanced triangles} 
 To validate social balance theory, we counted the triangles of each  {of the four possible types (see Fig. \ref{fig:socbalance_sketch})} 
 in each monthly signed social network  {obtained from the ALDM data}. Moreover, we compared the results in each case 
to a null model in which the signs were reshuffled among the links, to check
whether the balanced triangles were over-represented and the unbalanced ones under-represented with respect to this null model.
}
 {\paragraph{Triadic closure events} In social networks, a group of three individuals $A, B, C$ such that $A$ and $C$ are both friends of $B$ are called a triad centered at $B$. The triad is \emph{open}, forming a "wedge", 
if the link from $A$ to $C$ is missing. The process of closing an open triad to create
a triangle is called \emph{triadic closure} and is a well-known mechanism of evolution of social networks \citep{Rapoport:1953, Granovetter:1973}.}
 {In signed networks with positive and negative links, 
there are three possible wedge types: $++$, $+-$ and $--$ \citep{thurner2018virtual}. To investigate the dynamic aspects of
social balance, we considered the triadic closure events 
between successive months, in which an open "wedge" in a month $t$ became a closed triangle in the following month $t+1$. We thus
counted, for each wedge type in month $t$, how
many became triangles of each of the
four possible types (Fig. \ref{fig:socbalance_sketch}) and thus whether
closing wedges became preferentially balanced or
unbalanced triangles \citep{Szell:2010}.
}
 {\paragraph{Cosine similarity measure}
The cosine similarity \citep{Singhal:2001} is a measure 
defined between two vectors. It is 
bounded between $-1$ and $+1$, taking a value of $1$ if the vectors are identical, a value of $-1$ if they are opposite, and $0$ if they are perpendicular. 
In the case of temporal networks, let us consider a node $i$ and two different months ${t_1}$ and ${t_2}$. 
We denote by $ w_{ij}^{(t_1)}$ and $w_{ij}^{(t_2)}$ the weights of the links between individual $i$ and its neighbours $j$ in months $t_1$ and $t_2$, respectively. 
The local cosine similarity between months $t_1$ and $t_2$ is then defined for the ego-network of node $i$ as:
\begin{equation}
    CS_{t_1,t_2}(i) = \frac{\sum_{j} w_{ij}^{(t_1)}w_{ij}^{(t_2)}}{\sqrt{\sum_{j} \left(w_{ij}^{(t_1)} \right)^2} \sqrt{\sum_{j} \left( w_{ij}^{(t_2)} \right)^2 }} \ .
\end{equation}
It is thus equal to $1$ if $i$ not only has the same neighbours at $t_1$ and $t_2$ but also divides its co-presence time between them in the exact same way. It is equal to $0$ if $i$ has disjoints sets of neighbours in months $t_1$ and $t_2$. Note that, if all links are positive, the cosine similarity is bounded between $0$ and $1$.
Overall, an individual whose ego-networks changes strongly between $t_1$ and $t_2$ will have a low cosine similarity, whereas individuals whose ego-network is similar in both months will be associated with a high cosine similarity. 
The cosine similarity values between months therefore follow the evolution of the (in)stability of the considered
ego-network over time and we used the average value over all individuals as a global measure of the network’s stability between two months.}

 {We calculated the cosine similarity for every individual and for every pair of months. For each pair of months, we computed the average of the obtained values
over all individuals present in both months, to obtain a first global characterization of the network rearrangements between two months.
We then considered the whole histograms of the values for all nodes, to characterize
the heterogeneity of the amount of changes observed in the ego-networks of different
individuals. Finally, we considered "trajectories" of change for each individual, by computing $CS_{t,t+1}(i)$ for all $t$, i.e.,
by following for each individual the amount  of change of its ego-network from one month to the next.}

\subsection*{Behavioural data}

\paragraph{Observations}
We used the behavioural observations described by  {\cite{Claidiere2017}} and recorded between July $1^{st}$ and July $29^{th}$, 2014. 
During that time, the group included $22$ individuals, 7 males (mean age = 62 months, SD = 33) and 15 females (mean age = 124 months, SD = 75) ranging from 24 to 226 months. 
Observations were carried out by four trained observers using scan-sampling \citep{altmann1974observational}.
 {The data contained  $210$ behavioural observations per monkey per day during the study period,   {for} a total of 79 380 observations for the $22$ individuals.} For our study, we decided to focus solely on grooming as it is known to be a bonding activity in primates and therefore to represent affiliative relationships accurately 
\citep{Seyfarth:1977}.  More details on the behavioural observations can be found in  {\citep{Claidiere2017}}.

\paragraph{Construction of the grooming network}
We used the number of grooming events 
observed between a pair of individuals during the observation month  {(July 2014)} to build a weighted and undirected network. 
This network was then compared to the proximity network based on ALDM data of the same period.  {This
network resulted in $216$ links. A fully connected network with the same number of nodes would have had $231$ links. 
The weights (i.e. number of grooming events per dyad) ranged from 1 to 778 and the average and the median were respectively $60.7$ and $23.0$.}

\section*{ANALYSIS AND RESULTS}

Our first objective was to  {use the ALDM data to} build a signed social network for the  {entire} group. The co-presence network however contained only positive events (proximity events), so that we
first transformed it by a comparison with a suitable null model. The definition of avoidance (negative) links
made it possible to then analyse social balance and triadic closure events. We finally  {focused} on the temporal evolution of the network using various network metrics,
and  {showed} how to study it at various scales using similarity measures.

\subsection*{From weighted co-presence networks to affiliation and avoidance networks}
\label{subs_null_model_binary_network}

\begin{figure}[htb]
\centering
\includegraphics[width=.9\textwidth]{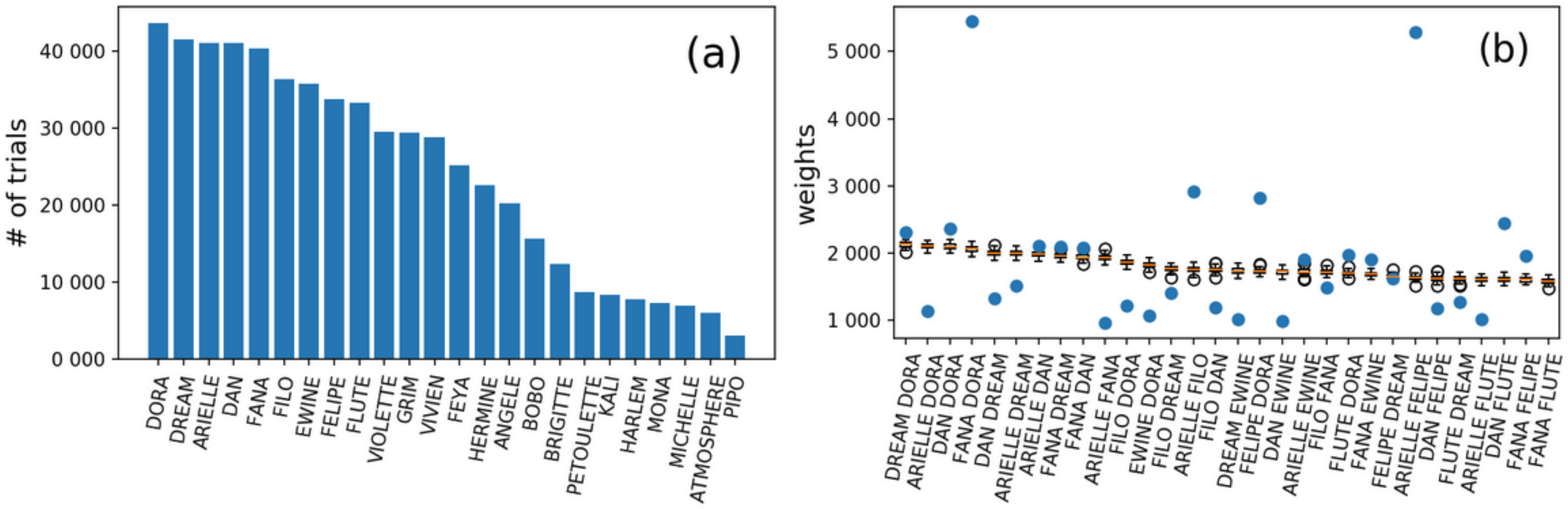}
\caption{
{\bf From the weighted co-presence network to the  signed social network.} 
a) Bar plot representing the number of cognitive tests performed by each individual in January 2014. 
b) For a sample of $30$ pairs of individuals, weights of the January 2014 co-presence network (blue dots) and boxplot showing the distributions of weights for the same pairs in networks resulting from the null model. In each box, the orange line marks the median and the extremities of the box correspond to the $25$ and $75$ percentiles; the whiskers give the $5$ and $95$ percentiles of each distribution.
 }
\label{fig:subfig_freq_boxplot}
\end{figure}

A first analysis of the ALDM data revealed that the baboon's frequency of participation in cognitive tests was very heterogeneous 
(see Fig. \ref{fig:subfig_freq_boxplot}a), as in  {\cite{Claidiere2017}}. This difference in participation could result in biased estimates of 
link strength  {in the
co-presence network defined in Methods,} for simple statistical reasons: individuals who are present more often in the ALDM booths
have a larger probability of being found in co-presence simply by chance,  
compared to individuals who participate less frequently. To take this behavioural heterogeneity into account and to determine which links 
could be interpreted as socially meaningful, 
we compared the observed co-presence to a null model based on the random permutation of the baboons' names 
for each month analysed,  {as described in the Methods section: this null model corresponds to the assumption 
that co-presence in the ALDM booths was independent from social relationships.} 

The results showed that some links had weights compatible with the null distribution, but 
others had observed weights that were above the $95\%$ confidence interval of the null distribution 
(see Fig. \ref{fig:subfig_freq_boxplot}b for illustrative examples), 
showing that the corresponding baboons were found in co-presence much more frequently than expected by 
 {chance, suggesting that they affiliated with each other.} 
Interestingly, some observed weights were significantly {\em lower} than expected by chance
(below the $5^{th}$ percentile of the null model distribution), 
showing that these baboons  {were in co-presence much less often than predicted by chance: 
it was then natural to assume that they were actively avoiding each other (see  \cite{Bejder:1998}). }
 {Note that pairs with no co-presence events, i.e. which were not linked in the original co-presence network, 
could be considered as avoiding each other if they were often found in co-presence in the null model.}

 {This comparison with a null model allowed us therefore to construct for each month
a new signed weighted network with both positive (affiliative) and negative (avoidance)
links. In this new network, the positive links were given by the 
co-presence links with weights above the $95^{th}$ percentile of the null model distribution.
The negative links on the other hand joined
the pairs of baboons with weights below the $5^{th}$ percentile of the null distribution.
Moreover, the weight of each link in the new network was given by the z-score value 
of the original weight with respect to the null model distribution (hence obtaining 
indeed {positive} values for the affiliative links and {negative} values for the avoidance links). Finally, for each month
we defined the set of positive links as the monthly affiliative network and the set of negative links as the monthly {avoidance} network.}

\subsection*{Validation of the ALDM network with the grooming network}
\label{subsec:validation_grooming}
To validate the network obtained through the ALDM system, 
we compared 
the network obtained  {from the ALDM data obtained during July 2014} to the one based on grooming behaviour (see Methods). 
Since grooming is an affiliative behaviour, we used  as a measure of similarity
the Pearson correlation coefficient between the weights of the affiliative network (i.e., using only the positive links) 
and the weights of the grooming network.  
 { {In addition}, as it is known that 
 the choice of the aggregation timescale can impact the characteristics of the network \citep{Ribeiro:2013}, we
 tested the robustness of the analysis by measuring the correlation between the affiliative and the grooming networks on various timescales, 
 ranging from $3$ to $25$ days
 within the same period (July 1-29, 2014). 
}

Note that, for timescales $T$ smaller than $14$ days, we could define
more than one observation period of such length  {within the period
of interest (July 1-29, 2014)}: we thus divided the total dataset in successive time windows of duration $T$ before building the affiliative and grooming networks, computing the correlation between their weights, and averaging over the time windows.

In addition, we compared the observed Pearson correlation to a  {null model} based on the random reshuffling of the baboon names in the grooming network. We simulated for each time window  {$1000$} such randomized networks,  {computed for each randomized network the correlation with the ALDM affiliative 
network, and built the distributions
of these Pearson correlation coefficients.} The results (Table \ref{table:pearsonr_grooming_copresence}) show that the 
observed Pearson correlation was well above the $95^{th}$ percentile  {of the random distribution} for all the timescales considered. 
 {We also observe that the correlation tends to increase for larger time windows, as 
also found by {\cite{Claidiere2017}}, since more observational data is included and thus
a more complete view of the social network is obtained.}

\begin{table}[]
\centering
\caption*{Correlation between grooming and co-presence networks}
\begin{tabular}{ccp{4cm}ccc}
\hline
Timescale (days) & Empirical value & Mean of the random distribution ($90\%$ CI) \\
\hline
3 & 0.257 & 0.065 (0.045-0.085) \\
5 & 0.309 & 0.066 (0.048-0.086) \\
7 & 0.309 & 0.069 (0.047-0.091)  \\
10 & 0.396 & 0.094 (0.076-0.113) \\
14 & 0.330 & 0.079 (0.053-0.111) \\
18 & 0.413 & 0.105 (0.087-0.122) \\
21 & 0.431 & 0.104 (0.085-0.123) \\
25 & 0.389 & 0.094 (0.065-0.124) \\
29 & 0.473 & 0.121 (0.103-0.138) \\
\hline
\end{tabular}
\caption{
Pearson correlation coefficients between the grooming and co-presence networks calculated for various aggregation timescales.
 {The last column gives the mean and $90\%$ confidence interval of Pearson
correlation coefficients computed
between the co-presence affiliative network
and $1000$ randomized versions of the grooming
network in which the baboon names were
randomly reshuffled.}
}
\label{table:pearsonr_grooming_copresence}
\end{table}

\begin{figure}[htb]
\centering
\includegraphics[width=.9\textwidth]{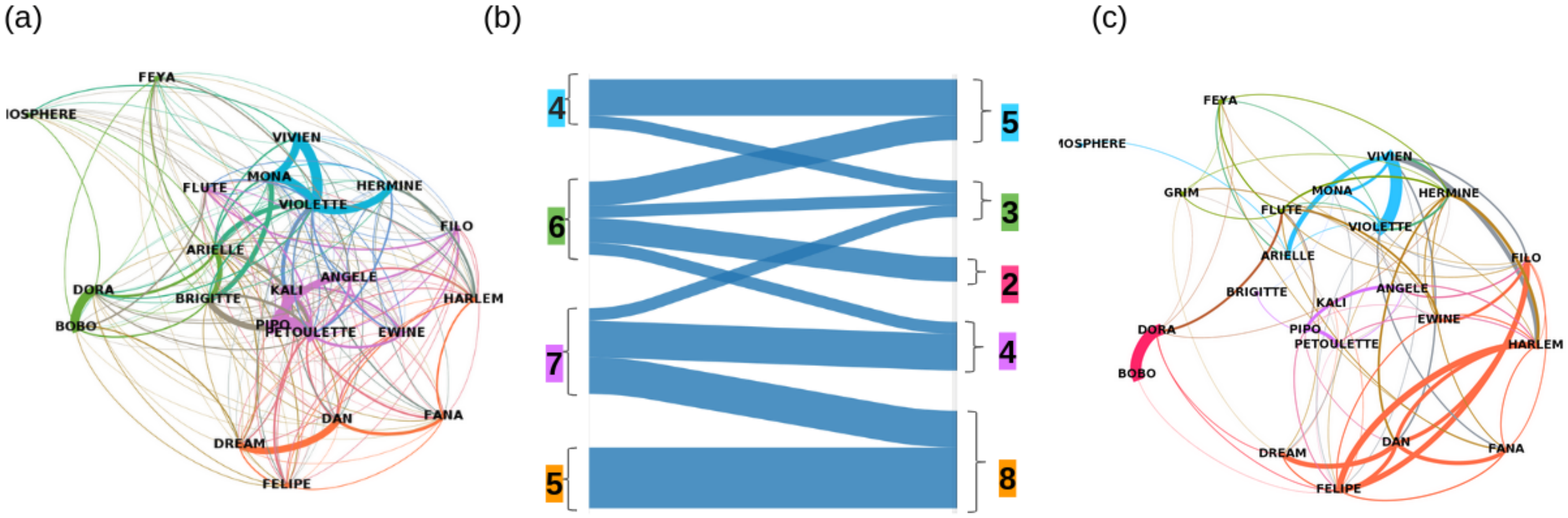}
\caption{
{\bf Affiliative network and grooming network.}
(a) and (c): Visualization, made using the Gephi software, of grooming (a) and affiliative (c) network of July 2014. The widths of the links are proportional to the weights of the networks (respectively number of grooming events and z-scores of the number of co-presence events), reflecting the strength of the relationships between nodes. Each colour correspond to a modularity class (i.e., a community) as assigned by the implementation (within Gephi) of the Louvain algorithm. The positions of the nodes were obtained by a Gephi layout implemented for the grooming network, and kept fixed for the affiliative network to facilitate visual comparison.
(b): Visualization of the differences in the community structures between the two networks
{by the flow of individuals across communities}. Numbers give the community sizes and the widths of the lines are proportional to the number of individuals common to grooming and affiliative networks communities.
}
\label{fig:subfig_groom_copres_net}
\end{figure}

The correlation between the grooming and ALDM networks is an overall measure of network similarity. 
Another well-known aspect of a  {social network}'s organisation, at an intermediate scale, is  {its} community structure
\citep{FORTUNATO201075}.
We determined the community structure of the affiliative and grooming networks using the Louvain algorithm \citep{Blondel:2008} 
implemented in the Gephi visualization software (www.gephi.org, see also \cite{bastian2009gephi}). 
To compare the resulting partitions (see Fig.
\ref{fig:subfig_groom_copres_net}), we calculated the adjusted Rand Index (ARI; \citep{rand1971, hubert1985}) implemented in the sklearn Python module \citep{pedregosa2011scikit}. The Rand Index takes values between $0$ and $1$, with $1$ indicating that the two partitions are the same and $0$ corresponding to a case in which the two partitions {disagree} on all pairs of elements. The ARI is a version of the Rand index corrected for chance, i.e., yielding a value close to $0$ when comparing two random partitions. Finally, we compared the observed ARI value with a distribution of ARI values obtained by randomizing the community structure of the grooming network, keeping the number and size of communities fixed but reshuffling individuals among communities. 
We observed an ARI value of $0.23$, well above the null distribution with mean $-0.0001$ (SD = $.067$), i.e., very close to $0$ as expected 
between random partitions.  

\subsection*{Social Balance: static and dynamic points of view}

Negative interactions such as avoidance behaviour are very difficult to observe  {among animals}. 
The construction described above however yielded a signed social network for each month, with both positive and negative links.
This made it possible to study social balance theory in the baboons'  {social network}.
To this aim, we counted in each monthly network the number of triangles of each type, i.e., with $0$, $1$, $2$ or $3$ negative links. 
 These numbers 
and also the total number of triangles fluctuated between months, but in all months
the balanced triangles were more numerous than the unbalanced ones. This was true both within the strong and the weak
versions of the social balance theory (see Fig. \ref{fig:socbalance_sketch}).

\begin{figure}[htb]
\centering
\includegraphics[width=0.8\textwidth]{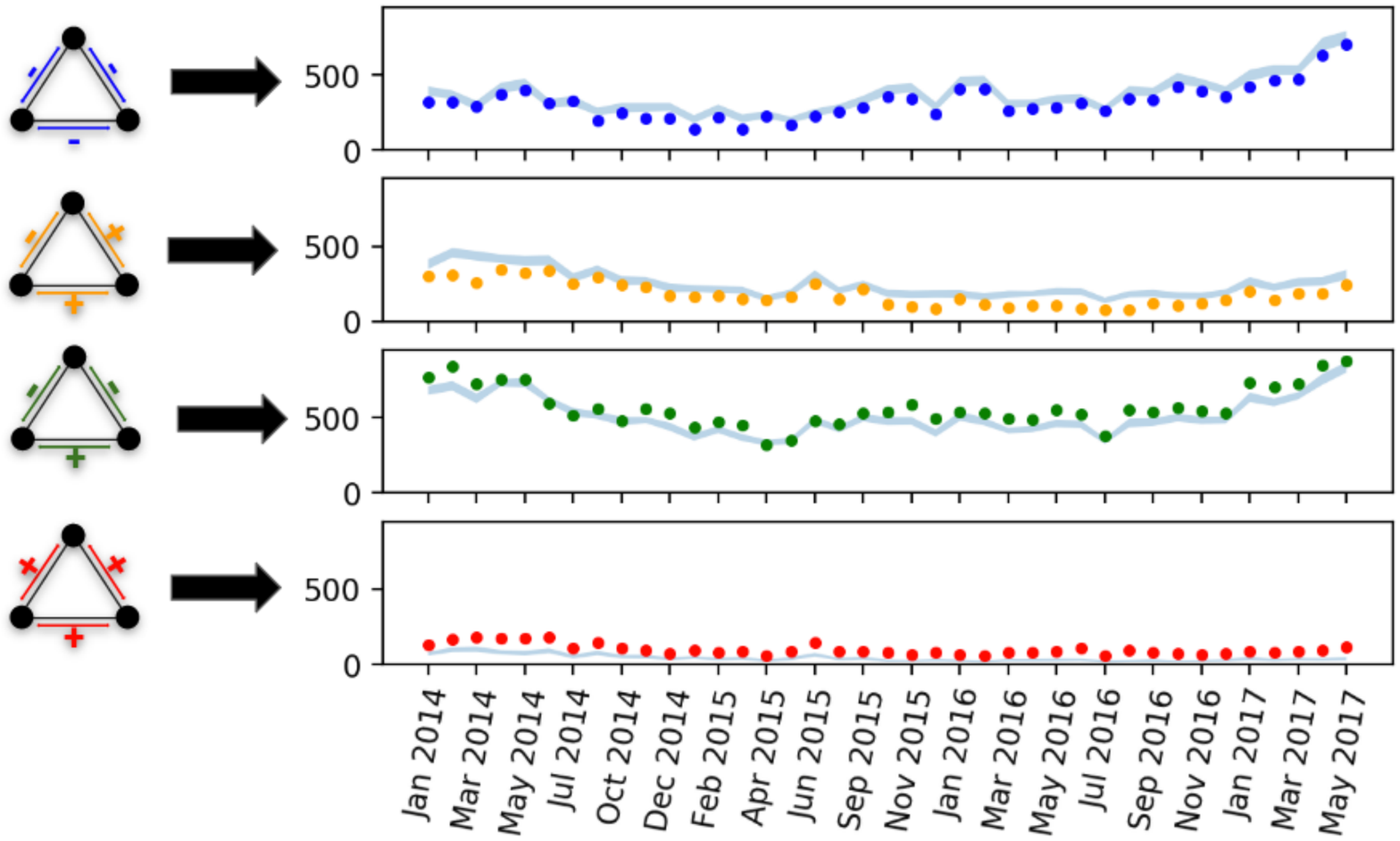}
\caption{\label{fig:triangles} 
{\bf Evolution of the number of signed triangles through time.}
Coloured filled circles: Number of triangles of each type (shown on the left part of the figure) in the empirical monthly signed co-presence networks (union of affiliative and avoidance networks). Shadowed area: confidence interval ($5^{th}$ to $95^{th}$ percentiles) of the distributions of numbers of triangles of each type in the randomized monthly networks. In the strong version of social balance theory, the triangle types in the two bottom panels are balanced, while the ones in the two top panels are unbalanced. In the weak version, triangles with three negative links (top panels) are also considered balanced.}
\end{figure}

Moreover, we compared the obtained numbers in each month with a null model in which the monthly 
network structure was fixed but the signs of the links were shuffled. Fig. \ref{fig:triangles} shows that
the triangles with three positive links were strongly over-represented in each month  {(100\% of the months)}
with respect to this null model.  It also shows that 
the triangles with only one positive link (balanced in both weak and strong balance theory) were also over-represented in most months 
(74\%  {of the months}), while the unbalanced triangles with two positive and one negative links were clearly under-represented (95\%  {of the months}). 
Triangles with three negative links tended to be moderately under-represented (82\%  {of the months}). We recall that  
these triangles are unbalanced in the strong version but balanced in the weak version of the social balance theory. 
Overall, balanced triangles were over-represented and unbalanced ones were under-represented with respect to the null model, 
showing that the signed monthly networks built from co-presence data did respect social balance theory.

\begin{figure}[htb]
\centering
\includegraphics[width=.9\textwidth]{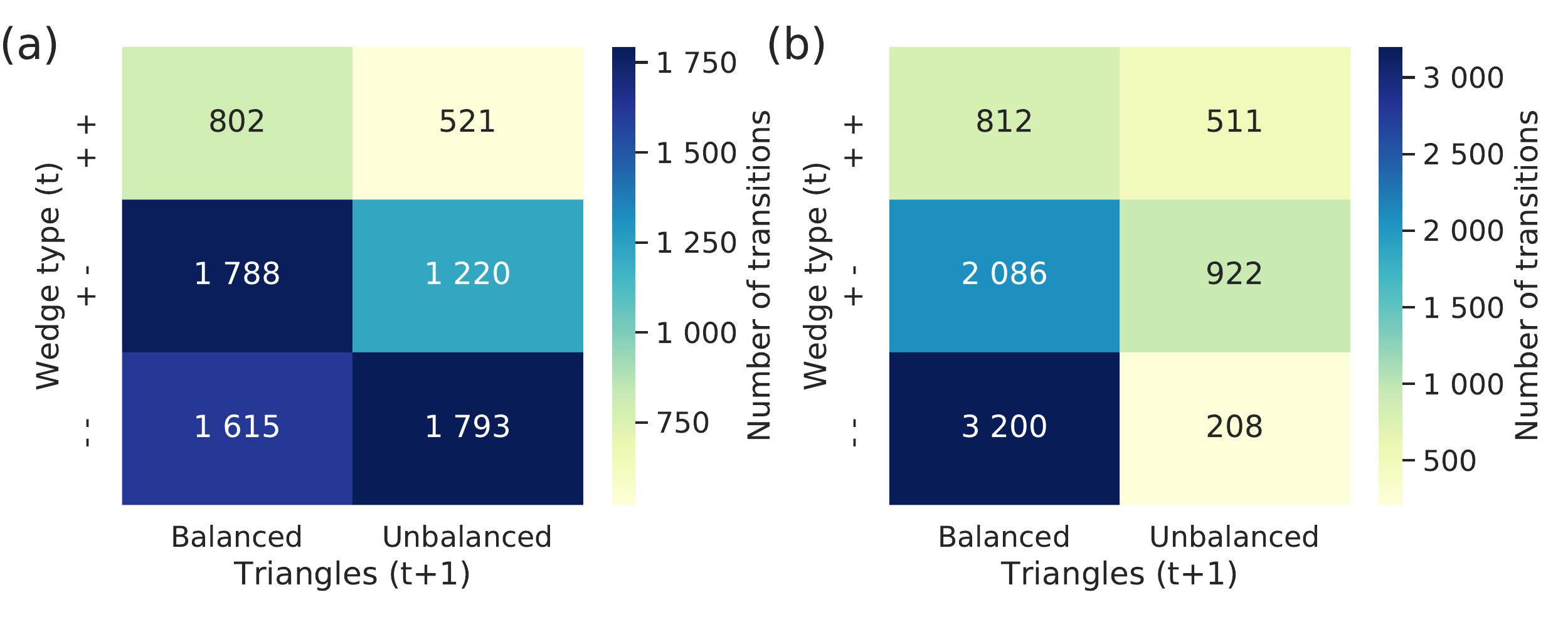} 
\caption{{\bf Social balance in the signed monthly networks.}
Total numbers of triadic closures of each type from a month $t$ to the next one $t+1$, i.e., 
numbers of transitions from the various types of wedges (from bottom to top,  $- -$, $+ -$, $++$) to balanced
triangles (left column of each table) or  unbalanced triangles (right column of each table). We present the numbers
of transitions summed over all the period of investigation (39 months) and for both the (a) strong and (b) weak 
formulations of social balance. For instance, over the whole period there were $3 200$ transitions from a wedge $--$
to balanced triangles, and $208$ to unbalanced ones, in the weak social balance formulation.}
\label{fig:triadic_closures_heatmap}
\end{figure}

In the union of the affiliative and avoidance networks, 
social balance theory implies that wedges 
(structures of two links $AB$ and $AC$ such that the link $BC$ does not exist, see Methods)
should preferentially close by forming balanced triangles. We thus calculated, for each pair of 
successive months $(t,t+1)$, the number of wedges of each type at $t$ 
($++$ with two positive links, $+-$ with one positive and one negative, and $--$ with two negative links) 
that became closed triangles at $t+1$ \citep{thurner2018virtual}. The results, summed over all values of $t$, 
showed that the total number of triadic closure events producing balanced triangles was larger than the total 
number of events producing unbalanced triangles, both using the strong or the weak version of the social balance 
(see Fig. \ref{fig:triadic_closures_heatmap}). Note that we obtained a substantial number of events in which a $--$ wedge 
became a $---$ triangle, which is considered balanced only in the weak formulation of social balance theory.

\subsection*{Stability and instability from network dynamics}

The long-term nature of the data set makes it an ideal setting to study the evolution of the social network structure and to investigate tools able
to detect periods of stability and instability. We first investigated the temporal evolution of the number of balanced and unbalanced triangles and
the number of triadic closure events of each type.  {Indeed, given the interpretation of the 
social balance theory that unbalanced triangles are a sign of tension and social stress, one could expect (i) an increase in the number of balanced triangles and a decrease in the number of unbalanced ones during periods of stability, and (ii) that a large number of unbalanced triangles could lead to an instability of the network and thus to important rearrangements 
in the following month.}
However, these numbers  
fluctuated widely from one month to the next (see  Fig. A\ref{fig:triadic_closures} in the Appendix), and
these variations showed no clear temporal signal or trend.

\begin{figure}[htb]
\centering
\includegraphics[width=\textwidth]{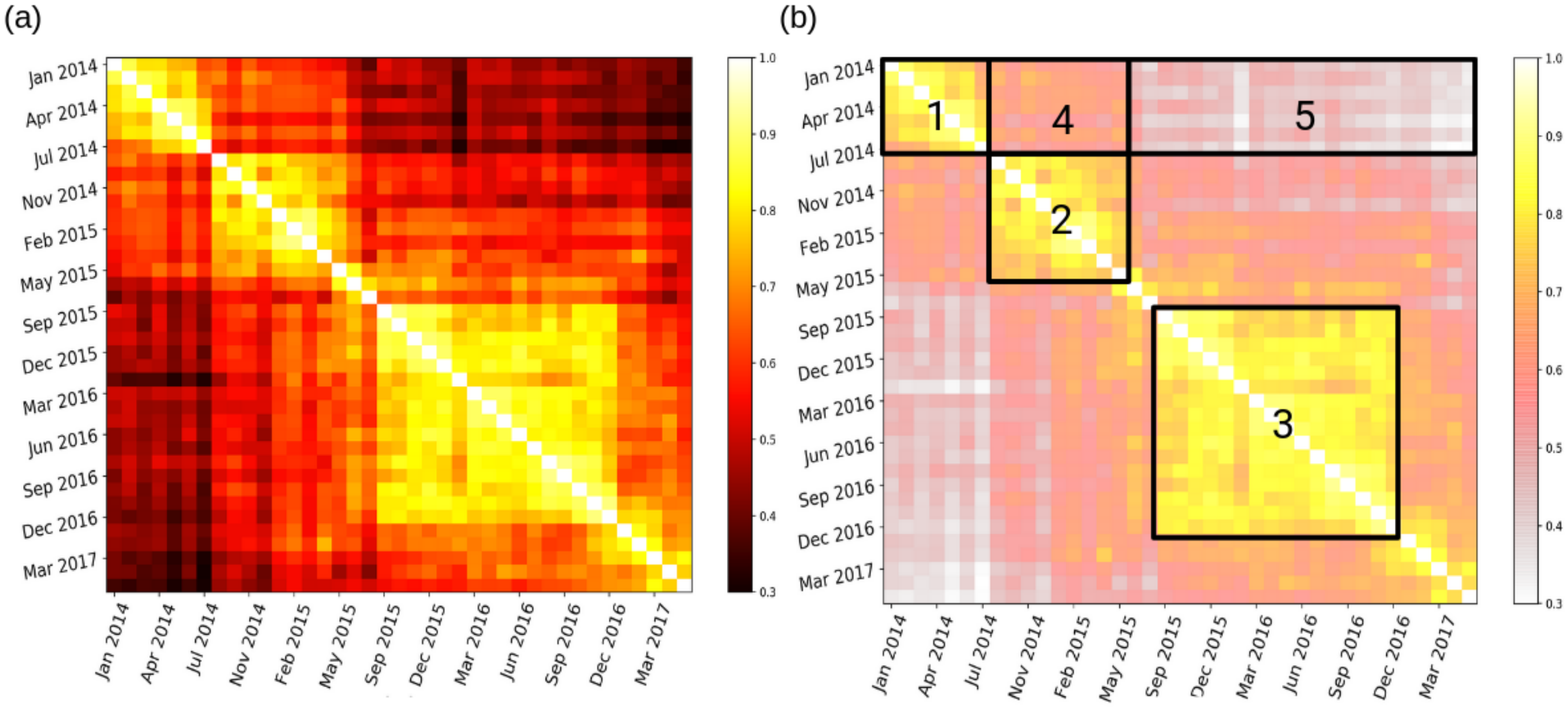}
\caption{\label{fig:matrices}
{\bf Dynamics of the social network at the
group level.}
(a) Colour-coded matrix of the average cosine similarity values for all pairs of months. Several periods of strong structural stability clearly appear as blocks of lighter (yellow-white) colour.
 {In contrast, the average CS between July and September 2014 was $0.577$, and the
one between July and and September 2015 was $0.65$. (b) the same matrix is represented with
highlighted periods of particular interest. The principal periods of stability are shown as block 1 
(January-July 2014: the average of the average CS values, excluding the 
diagonal, was $0.794$, SD $ = 0.07$), block 2 (September 2014-May 2015: average
of values $0.784$, SD $ = 0.054$), and block 3 (September 2015-December 2016: average of values $0.814$, SD $= 0.044$)}. 
 {Note that values tended to be progressively smaller away from the diagonal, i.e.
for months separated by longer and longer times. For example, the block 4, which represents the average CS values between the periods 
January-July 2014 and September 2014-May 2015, had a mean of $0.61$ and a SD $= 0.045$, while in block 5, between January-July 2014 and September 2015-December 2016, 
the mean value was $0.454$ and a SD $ = 0.047$.}}
\label{fig:cosin_sim_whole}
\end{figure}

We therefore decided to conduct a different investigation considering only the affiliative links.
To detect (in)stability in the resulting
affiliative network, we measured the similarity between monthly networks using the cosine similarity measure (see Methods), which we calculated for every individual and for every pair of months. 
 {The cosine similarity values at the group level (averaged over all individuals)
are shown in Fig. \ref{fig:matrices}a. The figure clearly shows that the average cosine 
similarity between different months remained very high in certain periods (yellow blocks along the diagonal). For instance, 
the average of the values obtained between two different months between 
January and July 2014 was $0.794$ (SD $= 0.07$). The average of the
values over different months taken between September 2014 and May 2015 was
$0.784$ (SD $= 0.054$). These examples are shown as blocks 1 and 2 in Fig.  \ref{fig:matrices}b.
Such large values of the average cosine similarity between different months imply that the ego-networks of the individuals did not change much, and therefore highlight periods of high network stability.}

 {On the other hand, at some moments the average cosine similarity between successive
months was lower, such as between July and September 2014 ($CS = 0.577$) and between July and September 2015 ($CS = 0.65 $).
These lower values indicate that the networks in those cases differed between the 
successive months, i.e., that some social network rearrangements took place, suggesting potential periods of network instability. }

We also note that the average cosine similarity values tended to be progressively lower away from the diagonal, i.e.
for months separated by longer and longer times. For instance, the block 4 in Fig.  \ref{fig:matrices}b corresponds
to all pairs of months ($t_1$,$t_2$) with $t_1$ between January and July 2014 and $t_2$ between September 2014 and May 2015: 
 the average of the values within this  block was $0.61$ (SD $= 0.045$). When considering instead block 5 in  Fig. \ref{fig:matrices}b, which
 correspond to comparing a month in the period January-July 2014 to a month between September 2015 and December 2016,
 the average of the cosine similarity values was  $0.454$ (SD $= 0.047$). This is consistent with the fact that the social network 
 became more and more different, and did not come back to a previous structure during the period of study.

\begin{figure}[htb]
\centering
\includegraphics[width=\textwidth]{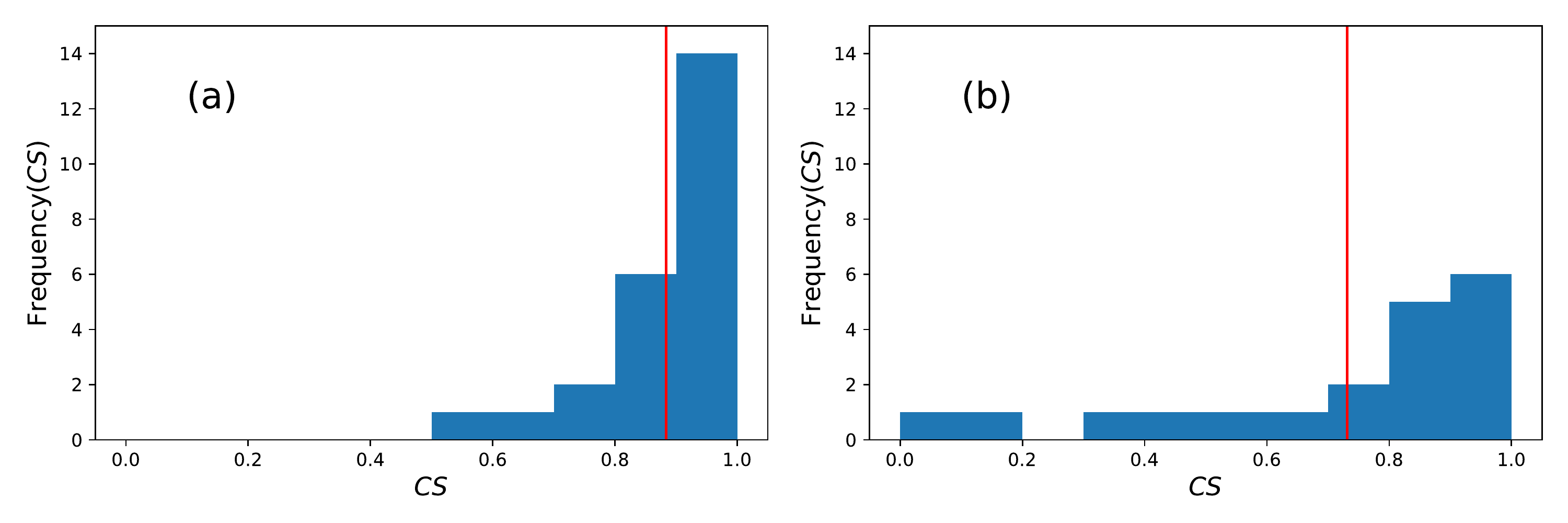}
\caption{\label{fig:distr_cosin_sim_4_5_16_17}
{\bf Ego-network dynamics.}
Histograms of individual's cosine similarity values between weighted ego-networks for two different pairs of successive 
months: (a) April 2014, May 2014; (b) April 2015, May 2015.  {For each panel, the histogram corresponds to the cosine similarity values 
of all the ego-networks of the individuals present in both months. The vertical line gives the average value of the cosine similarity (i.e., the value shown
in Fig. \ref{fig:matrices}a), that correspond to $0.884$ and $0.731$ for April 2014, May 2014 and for April 2015, May 2015 respectively.}}
\label{fig:hist_cos_sim}
\end{figure}

Fig. \ref{fig:distr_cosin_sim_4_5_16_17} sheds more light into these two types of periods by presenting the distributions of individual cosine similarity values for two periods corresponding to high and low group level similarity. 
When the similarity is high (Fig. \ref{fig:distr_cosin_sim_4_5_16_17}a), the distribution is highly skewed with most individual's cosine similarity values close to $1$. This shows that the network global stability was the result of the stability at the individual's level. In contrast, when the
average similarity is lower (Fig \ref{fig:distr_cosin_sim_4_5_16_17}b), the distribution of the individual cosine similarity values is broader with some individuals 
maintaining their ego-network (large value of the cosine similarity) while others changed dramatically (cosine similarity values close to 0). 
This implies that the instability of the network was not a global one but that some parts of the networks remained stable while others underwent important changes.

\begin{figure}[htb]
\centering
\includegraphics[width=\textwidth]{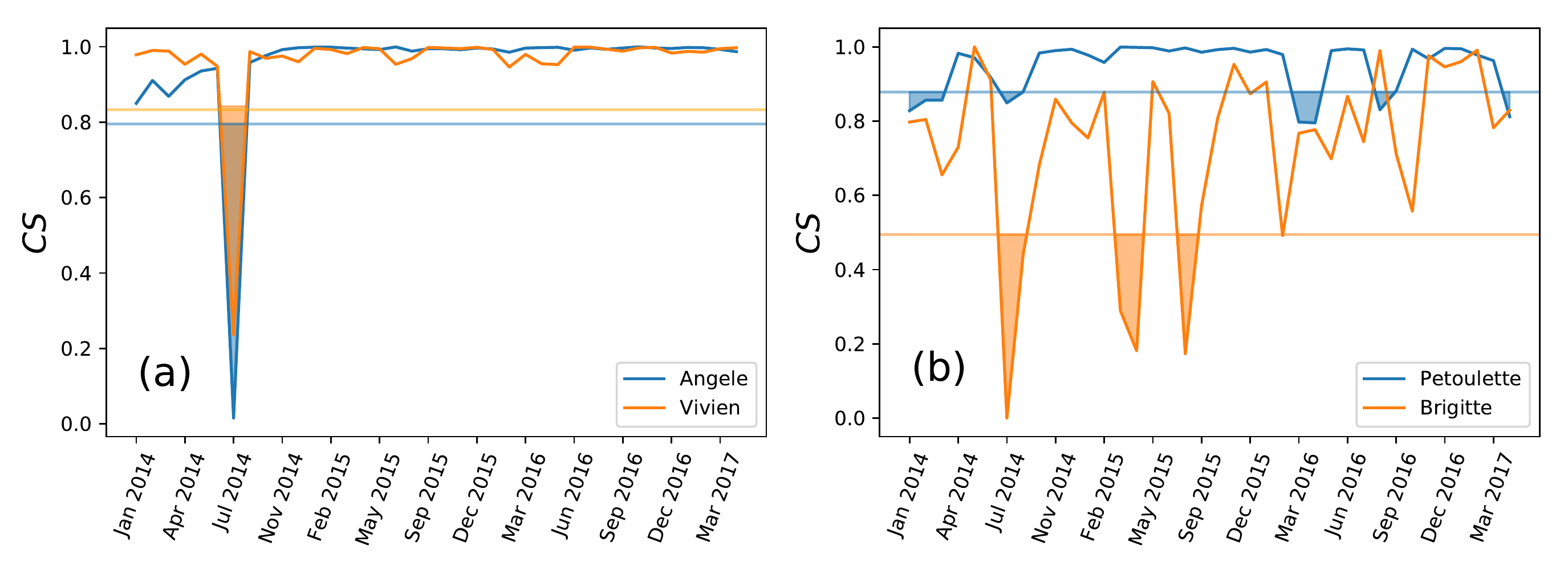}
\caption{{\bf Dynamics of ego-networks.}
Evolution of ego-network cosine similarity values computed between one month and the next
for several individuals ($CS_i(t,t+1)$ for each month $t$).  {The lines give for each individual
the average of its cosine similarity value over time, minus one standard deviation. The filled areas correspond therefore to 
very unstable periods, i.e. to periods where the ego-network cosine similarity between successive months are more 
than a standard deviation below the average.} The two individuals in panel (a) showed a 
stable ego-network overall  {(Vivien: mean = $0.96$, SD = $0.12$; Angele: mean = $0.95$, SD = $0.15$)} but displayed 
a sudden and synchronized change 
between July and September 2014. The two individuals in panel 
(b) had very different patterns of ego-network stability and instability  {both in terms of variability of values (Petoulette: mean = $0.94$, SD = $0.12$; Brigitte: mean = $0.71$, SD = $0.06$) and in terms of (absence of) synchronization: 
for instance in August 2016 we note that while Brigitte's trajectory has a local maximum ($0.93$), Petoulette's trajectory undergo a local minimum ($0.83$)}.
}
\label{subfig:ind_traj_synchro_no_synchro}
\end{figure}

In order to better understand the  {network's} dynamic, we studied the evolution of ego-networks independently. Fig. \ref{subfig:ind_traj_synchro_no_synchro} shows
the cosine similarity between the ego-networks of specific individuals in successive months. 
The resulting patterns differed greatly between individuals (we show the results for all individuals in the Appendix). 
For instance,  {as can be seen in Fig \ref{subfig:ind_traj_synchro_no_synchro}a,} Vivien (adult male) and Angele (adult female) had on average high cosine similarity values over the entire study (resp. $0.96$ and $0.95$), showing prevalently a strong stability of their ego-networks, but these averages hide an interesting pattern: in a synchronized way, these two baboons went through a strong rearrangement of their ego-networks, with cosine similarity values of respectively $0.23$ and $0.01$, between July and September $2014$. Both individuals had thus very stable 
ego-networks before and after this period, but their ego-networks between the first and second stability periods differed strongly. 
 {In particular, between these two periods Angele lost her strong link with Pipo  {(adult male)}, the links between Vivien and
most of his females became much weaker, and a strong link between Angele and Vivien appeared.}
On the other hand, both individuals kept a stable ego-network during other structural changes observed in the matrix of Fig. \ref{fig:matrices}a,  {such as 
between March and June 2015: this
highlights once again, as deduced
also from Fig. \ref{fig:distr_cosin_sim_4_5_16_17}b, that
a low average cosine similarity between two
different months can be the result of
a large variation of some ego-networks while
others remain completely unchanged}. 
 {Fig. \ref{subfig:ind_traj_synchro_no_synchro}b
shows that other individuals went through very different patterns of ego-network stability and instability: some kept a relatively
stable ego-network throughout the whole
period, with only small changes (for instance, Petoulette,  {adult female}, in 
Fig. \ref{subfig:ind_traj_synchro_no_synchro}b),
and some had much more important and frequent
changes in their local ego-network
(such as Brigitte,  {adult female}). Fig. A\ref{fig:ind_trajectories_red_emph_dt_5}
in the Appendix displays the evolution
of the ego-network cosine similarity values for all the individuals.
}

\begin{figure}
\centering
\includegraphics[width=0.9\textwidth]{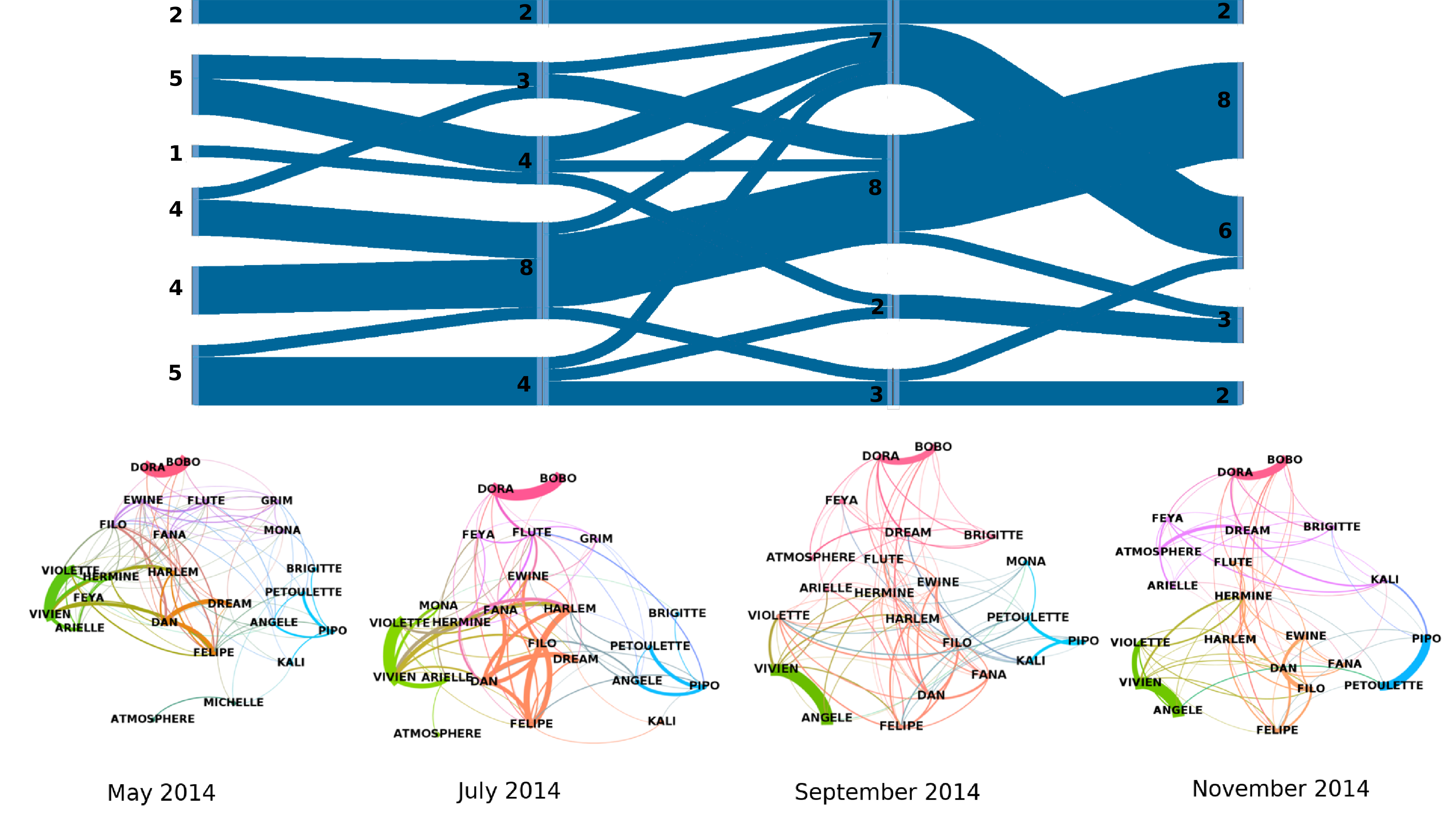}
\caption{\label{fig:flow_graphs_4_months} 
{\bf Visualization of the co-presence networks and of the flow between communities across the months of May, July, September and November 2014. }
Different colours in the graphs correspond to different communities. In he flow diagram, the numbers specify the community size and the widths of the lines are proportional to the number of individuals common to the communities joined by the lines.}
\end{figure}

Finally,  {we sought to analyse in more details the important} change in the network between July and September 2014 (see Fig. \ref{fig:matrices}a).
 {Indeed, the} average cosine similarity between these months was equal to $0.577$, 
while the average between May and July was
$0.779$, and $0.845$ between 
September and November 2014. As already discussed above, these average values hide a heterogeneous amount of local rearrangements, 
as shown by the distributions of Fig. \ref{fig:distr_cosin_sim_4_5_16_17}. Fig. \ref{fig:flow_graphs_4_months} shows a visualization of the affiliative networks in these months, as well as the flux of individuals between communities obtained by the Louvain algorithm. The ARI between the resulting partitions in July and September was 
equal to $0.16$, while it was equal to $0.67$ between the partitions in communities of the September and November networks and to $0.41$ between partitions in communities of May and July networks. 

The visualization reveals several important changes in the network:
\begin{itemize}
    \item Angele moved from a community to another. The strong link with Pipo disappeared and a new strong link with Vivien emerged; 
    \item Links between Vivien and other females of his community weakened as the link with Angele abruptly became very strong;
    \item The number of individuals went from 23 to 22 because Grim was removed from the group for a medical reason unrelated to the experiment.
\end{itemize}
The two first points are clearly related to the synchronization in the timelines of Fig. \ref{subfig:ind_traj_synchro_no_synchro}a. The visualization also made clear how certain parts of the network were on the contrary quite stable between these months, as discussed above in relation to Fig \ref{fig:distr_cosin_sim_4_5_16_17}b.

Note that  {all these points were deduced
only} from the study of the affiliative network built from the proximity in ALDM booths,  {without any external knowledge}. 
Interestingly, it turned out that direct annotations of observations of the group in the period of the summer 2014 confirmed that 
 {two important changes occurred in July 2014, namely: (i) Grim died and  (ii) Angele changed primary male during that month.}

\section*{DISCUSSION}

In the present study, we have analysed one of the largest (more than 16 million records) long-term (more than 3 years) high resolution dataset collected on a group of non-human primates. We then used this dataset to test social balance theory and to study the dynamic patterns of  {social relationships'} (in)stability in a non-human primate.

We used this dataset, composed of automatically collected cognitive tests performed by a group of Guinea baboons, to build a co-presence network. By comparing the observed co-presence to a random distribution generated under the assumption that  {the co-presence in neighbouring
ALDM booths was due only to randomness and independent from the actual} social relationships among the baboons \citep{Manly:1997,Bejder:1998,Farine:2017}, we were able to establish a signed network representing both affiliative and avoidance relationships. This network building procedure has two main advantages. First, it takes into account the heterogeneity in the number of records of different individuals, thus reducing potential biases in the estimation of bond strength. Heterogeneity in the number of observations of individuals is a common feature of social network analysis in animals and several association coefficients have been developed to limit common biases (see e.g. \cite{Whitehead:2008}). Developing a null-model of social interaction to estimate link strength however, has the second advantage of allowing the researcher to determine both affiliative and avoidance relationships from positive interactions simply by  {assuming that individuals who meet more than chance are actively seeking each other's company while individuals who meet less than chance are actively avoiding each other
\citep{Bejder:1998}}. Notably, negative interactions are more difficult to study in nature than positive interactions because they are often less evident (such as avoidance) and/or less frequent (such as open fights). Consequently, little attention has been paid to agonistic social networks but according to social balance theory for instance, negative links are crucial to understand the social evolution of a group. Using the lack of positive interactions, compared to a null-model, as an indicator of a negative relationship could therefore prove a useful general tool. 

Indeed, in our study we were able to use the signed network to show that the networks followed the predictions of social balance theory \citep{heider1946attitudes}. From a static point of view, the results showed that balanced triangles were over-represented and unbalanced triangles under-represented when compared to a null distribution based on random permutation of the networks’ edge signs. Furthermore, from a dynamic point of view, we found that wedges (unclosed triangles) tended to close into balanced triangles more frequently than into unbalanced ones. Interestingly, we also observed many closure events towards triangles with three negative edges: this 
could be linked to the fact that the group of baboons was composed of more than two sub-units, so that triangles with three avoidance links were not rare. 

Our results are in line with two previous studies that aimed at testing social balance theory directly. \cite{Szell:2010} in an online game with more than 300 000 participants and \cite{ilany2013structural} in a study of rock hyraxes, both found that their networks were generally balanced with an over-representation of balanced triangles (especially $+++$ triangles) and an under-representation of unbalanced triangles (especially $++-$ triangles). In these two studies, as in ours, there was substantially more support for the weak formulation of social balance theory  \citep{davis1967clustering}, compared to its stronger alternative. Finally, \cite{Szell:2010} also studied the triangle closure dynamic and also found, like us, that more wedges (unclosed triangles) closed into balanced triangles than unbalanced ones. Based on these three studies with different species and in very different contexts, it is tempting to conclude that (weak) social balance theory may represent general principles of social organisation, as envisaged by \cite{heider1946attitudes}. Surprisingly however, very few studies have tested social balance theory and in our opinion it deserves more scrutiny especially because it provides a clear static and dynamic theoretical framework to understand the structure of social networks and their evolution.

Unfortunately, the analysis of the temporal evolution of social balance measures (in proportion or number of balanced vs. unbalanced triangles with both the weak or strong interpretation) did not yield clear insights into the temporal evolution of the  {structure of the social network} (see Appendix). There were no clear changes in the different measures between periods of social stability and instability  {that were instead revealed by the analysis of the cosine similarity}. 
One explanation could be that our analysis did not consider the correct timescale on which changes in social balance occur. For instance, there is some evidence in chacma baboons (\textit{Papio ursinus}) that individuals adapt their social strategies on very different timescales
 \citep{Sick:2014,Henzi:2009,Silk:2010}. If the social network we studied can be reorganised rapidly, going from balanced to unbalanced to balanced again on a scale of a few days, then we should expect to  
find a large difference in a similarity measure (the network has changed) but little difference in terms of the number of balanced-unbalanced triangles on a monthly timescale. This problem may not be easy to solve because networks established on shorter timescales are subject to imprecise estimates of link strengths, which in turn could mask changes in the network. In our study for instance, we found that the monthly timescale usually recommended 
\citep{Whitehead:2008}, represented a good compromise between a precise estimate of link strength and the possibility to detect temporal changes. 
However, the best aggregation timescale is likely to change depending on the situation, study species, group and data gathering technique 
\citep{Farine:2015,Davis:2018}.  {The 
investigation of the cosine similarity measure also shows a limit of the social balance theory. We observed important changes in the
structure of the network (Fig. \ref{fig:cosin_sim_whole}) that were not reflected by changes in the social balance of the social network,
probably because they were linked to changes
occurring in the relationships of only
some of the individuals
(Fig. \ref{fig:hist_cos_sim}). This suggests that the measure of social balance through
the counting of balanced and unbalanced
triangles may be too coarse and too global
a measure to be able to detect and follow changes
in the relationships between individuals. In 
particular, social balance focuses on 
the sign of these links but does not take
into account their weights and hence the 
relative importance of the relationships in 
the social network.}

We were nevertheless
able to study the  {social network's} dynamics 
using the cosine similarity between the affiliative ego-networks in successive months. 
Cosine similarity is a very versatile measure
and can be used to compare networks over different timescales but also across scales within the network, from single individuals to sub-groups to the entire network. The group averaged cosine similarity value allowed us to visualize periods of global stability, and to identify moments of instability (Fig. \ref{fig:matrices}a).  
More detailed investigation of the cosine similarity values showed that the periods of stability corresponded to high stability (large similarity values) for almost all individuals, as could be expected.
However, the instability revealed by the average did not necessarily come from an instability of the entire network, but rather from a mixture of locally stable and changing structures (Fig. \ref{fig:distr_cosin_sim_4_5_16_17}). 
Following individual cosine similarities therefore allowed us to identify, for each period of interest, individuals with more or less stable ego-networks, as well as interesting patterns of synchronization of ego-networks evolution 
(Fig. \ref{subfig:ind_traj_synchro_no_synchro}). The patterns we identified on a monthly timescale often suggest sudden and important changes that correspond, at least in some cases, to adult females changing primary males, as observed in the wild 
\citep{Goffe:2016}. We suspect that stronger perturbations of the network could be linked to high ranking females changing males (such as Angele), whereas smaller perturbations could be linked to more peripheral females (such as Brigitte)  {changing principal male}.

 {Importantly, if some individuals  {in a wild population} have more stable ego-networks compared to other members of the group, 
we  {may} expect these individuals to have higher fitness, with increased longevity and a greater number of offspring for instance, because stable social relationships have been associated with all these factors, especially in baboons \citep{Silk:2003,Silk:2009,Silk:2010,Alberts2019}. However, this effect on fitness may be more difficult to study in captivity where food, health and reproduction are controlled.}

\section*{CONCLUSION AND PERSPECTIVES}
 Social interactions are an important fitness component of group living animals (see e.g. \cite{Alberts2019}) and social network analysis provides powerful tools to describe social interactions and analyse their evolution through time \citep{Hinde:1976}. In fact, social network analysis has transformed research in ecology and evolution (for a review see \cite{Cantor:2019}). Here, we have  
  {analysed} the temporal (in)stability of a social group of baboons using an automatically collected high-resolution long-term dataset. We have developed general tools to construct a signed network and shown that our study group’s social network respected the predictions of social balance theory. However, the use of a  
similarity measure proved to be more sensitive and more versatile to understand changes in the individuals' social relationships and their consequences at the group level. In particular, our results show that behind what, at first glance, looks like a stable social network, there is a complex and subtle mixture of stable and unstable ego-networks. In the future, long-term high-frequency data (see \cite{Krause:2013} for a review of recent technological developments) could help determine the fitness consequences of individuals' social strategies.

\section*{DATA AVAILABILITY}
The data analyzed in this article are freely available on the Open Science Foundation website, with the following DOI:10.17605/OSF.IO/NX2PJ

\section*{ACKNOWLEDGMENTS}

{The authors declare no competing interests. N.C. and J.F. gratefully acknowledge financial support from the ASCE program
(Grant No. ANR-13-PDOC-0004) of the Agence Nationale de la
Recherche.
The funders had no role in the study design, data collection
and analysis, decision to publish, or preparation of the manuscript. This
research was conducted at the Rousset-sur-Arc 
Primate Center (CNRS-
UPS846), France. The authors thank its staff for technical support and Julie Gullstrand for helping collecting the observational data.



\section*{References}

\bibliography{Biblio} 

\renewcommand{\thefigure}{\arabic{figure}}
\newcounter{tabcountSI}
\renewcommand{\thetabcountSI}{S\arabic{tabcountSI}}
\newcommand{\tabcountSI}{\refstepcounter{tabcountSI}}
\newcounter{seccountSI}
\renewcommand{\theseccountSI}{S\arabic{seccountSI}}
\newcommand{\seccountSI}{\refstepcounter{seccountSI}}
\clearpage
\newpage
\section*{Appendix}

\subsection*{Robustness of the analysis with respect to the ALDM aggregation time window}

In the Methods section, we described how we built a temporal co-presence network based on the temporal and spatial 
proximity of the baboons in the ALDM workstations: we aggregated the raw data in time windows 
of length $\Delta t = 5s$ and considered that the individuals performing tests in the same time window and in neighboring workstations
were in co-presence.

For the sake of completeness, we report here results obtained with a different value of the aggregating
window, namely $\Delta t = 10s$, to show the robustness of the observed phenomenology.

 Fig. \ref{fig:triangles_dt_10} and \ref{fig:triadic_closures_heatmap_dt_10} show 
the results concerning social balance theory, and  Fig. \ref{fig:avg_cosin_sim_dt_10}
displays the colour-coded matrix of the average individual cosine similarity values between
the monthly affiliative networks for all pairs of months. In all cases,
results very similar to the ones presented in the main text were obtained.

\begin{figure}[hb]
\centering
\includegraphics[width=0.7\textwidth]{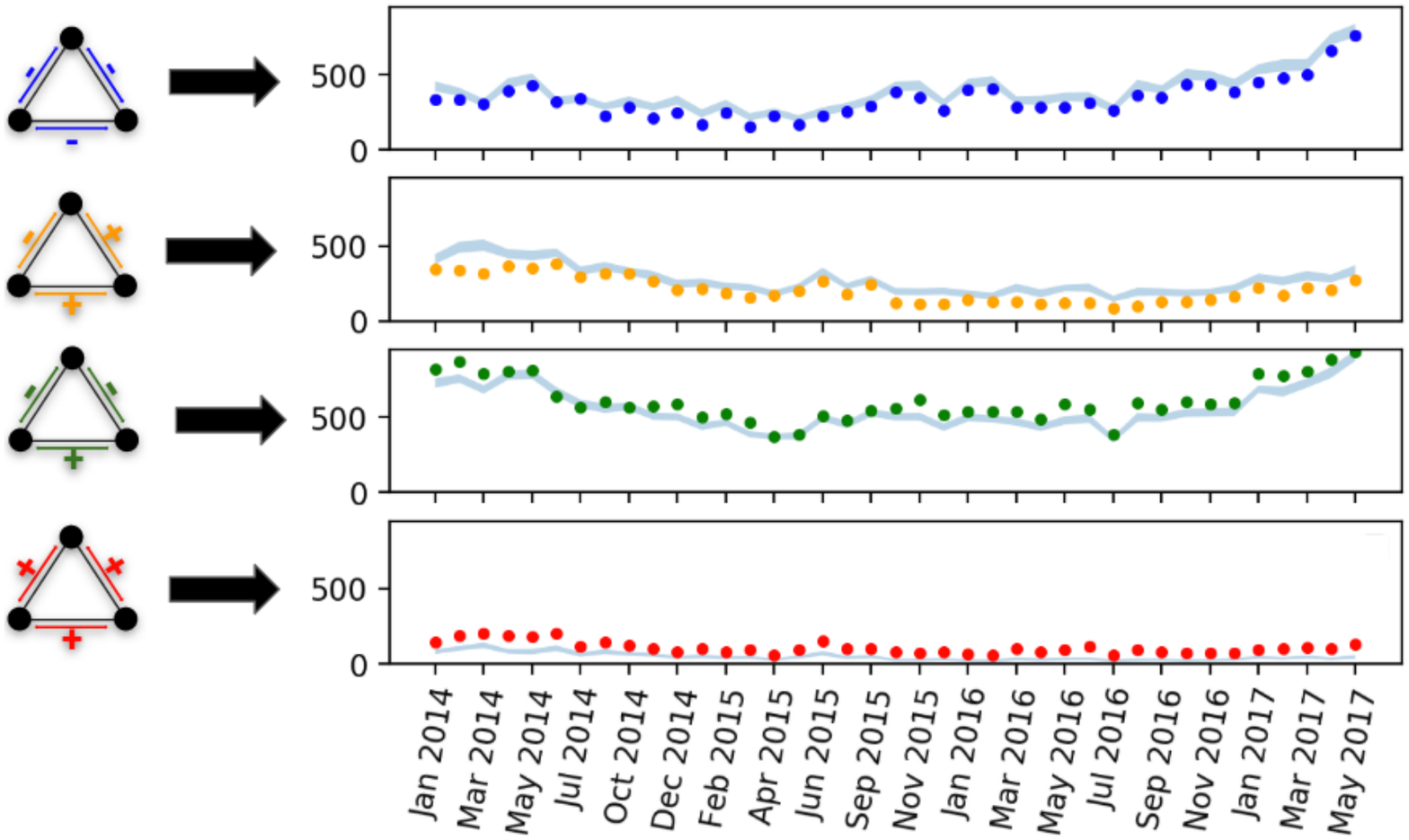}
\caption{\label{fig:triangles_dt_10}
{\bf Evolution of the number of signed triangles through time ($\Delta t = 10s$).}
Number of triangles of each type in the 39 monthly signed co-presence networks built using $\Delta t = 10s$.
As in the main text, the shadowed areas correspond to the confidence interval ($5^{th}$ to $95^{th}$ percentiles) 
of the distributions of the numbers of triangles of each type in randomized monthly networks.}
\end{figure}

\begin{figure}
\centering
\includegraphics[width=\textwidth]{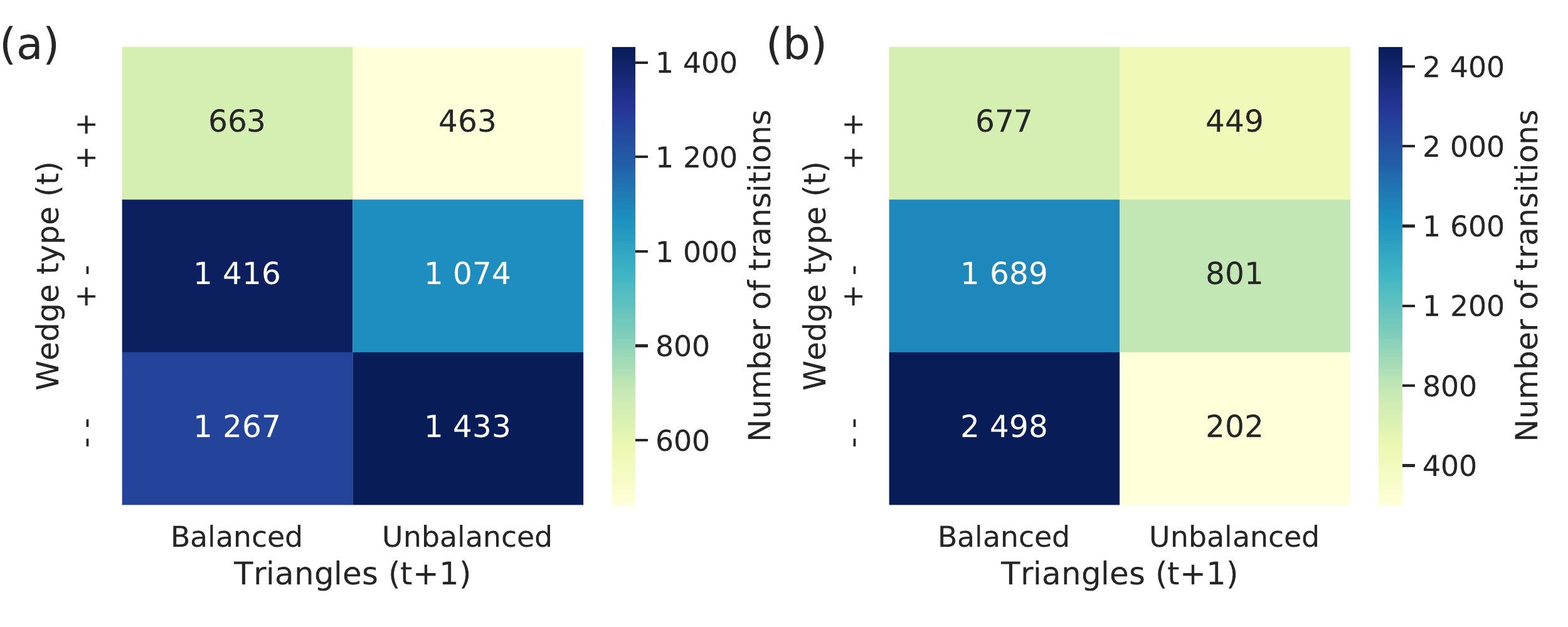}
\caption{{\bf Social balance theory from the 
dynamical point of view in the signed monthly networks built with $\Delta t = 10s$:}
Numbers of transitions from one month to the next, from the various types of wedges 
to balanced or unbalanced triangles, summed over all the period of investigation (39 months) 
and for both {strong} (a) and {weak} (b) formulations of social balance. }
\label{fig:triadic_closures_heatmap_dt_10}
\end{figure}

\begin{figure}
\centering
\includegraphics[width=0.7\textwidth]{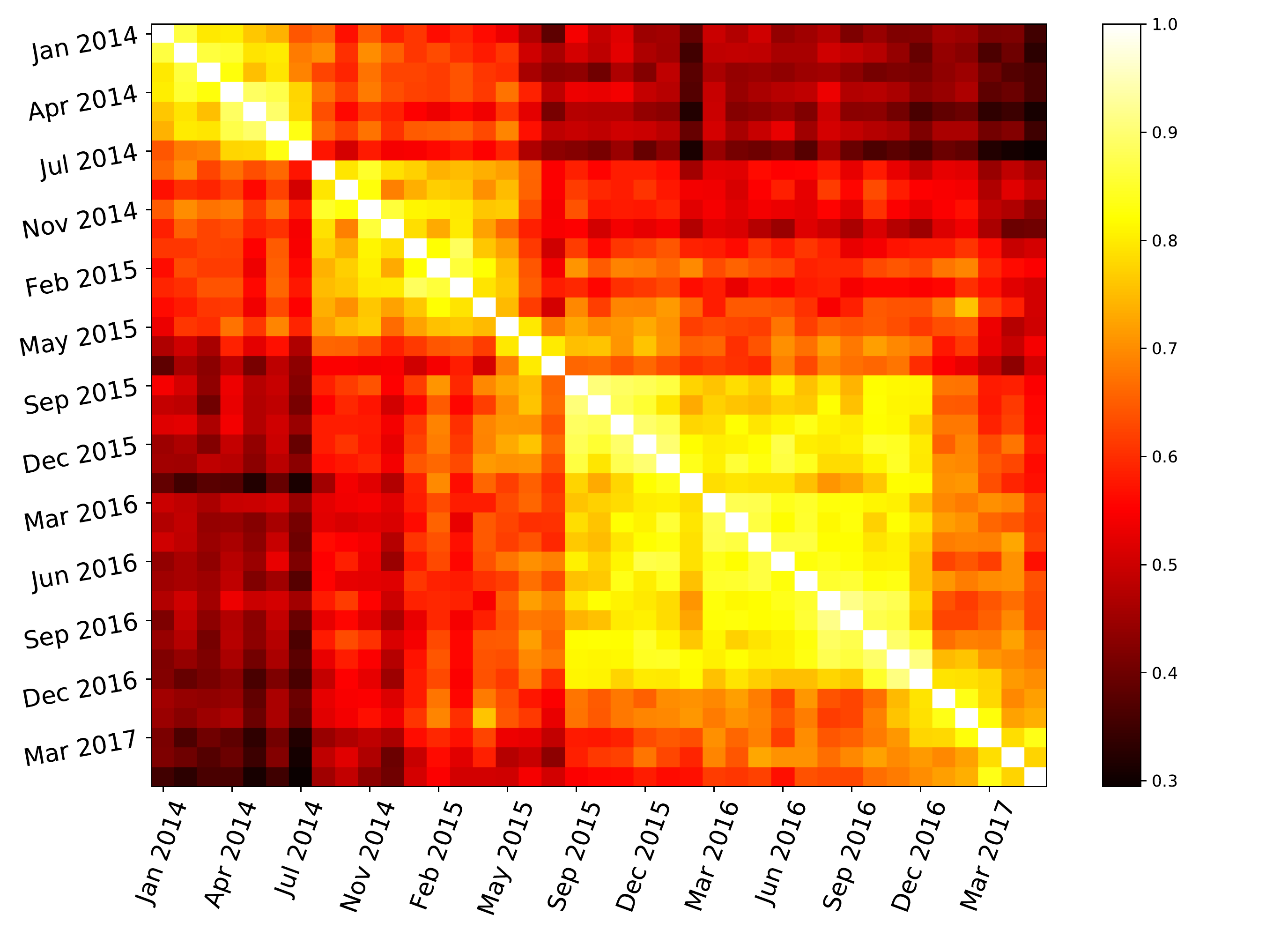}
\caption{\label{fig:avg_cosin_sim_dt_10} 
{\bf Group level dynamics for the monthly networks built using $\Delta t=10s$.}
Colour-coded matrix of the group cosine similarity values for all pairs of months. Patterns of stability and instability are the same as the case of $\Delta t = 5s$ shown in the main text. }
\end{figure}

\subsection*{Temporal evolution of triadic closure events and of indicators linked to the numbers of triangles}

Fig. 6 gives the global number of triadic closure events from one month to the next. These events are defined by the fact that a wedge in one month
(a structure of two links $AB$ and $AC$ such that the link $BC$ does not exist) becomes a triangle
in the next month. We classified these events depending on the type of wedge 
and on the type of resulting triangle (balanced or unbalanced). In 
Fig. \ref{fig:triadic_closures}, we show the numbers of each type of event for each month.

\begin{figure}[hb] 
\centering
\includegraphics[width=.85\textwidth]{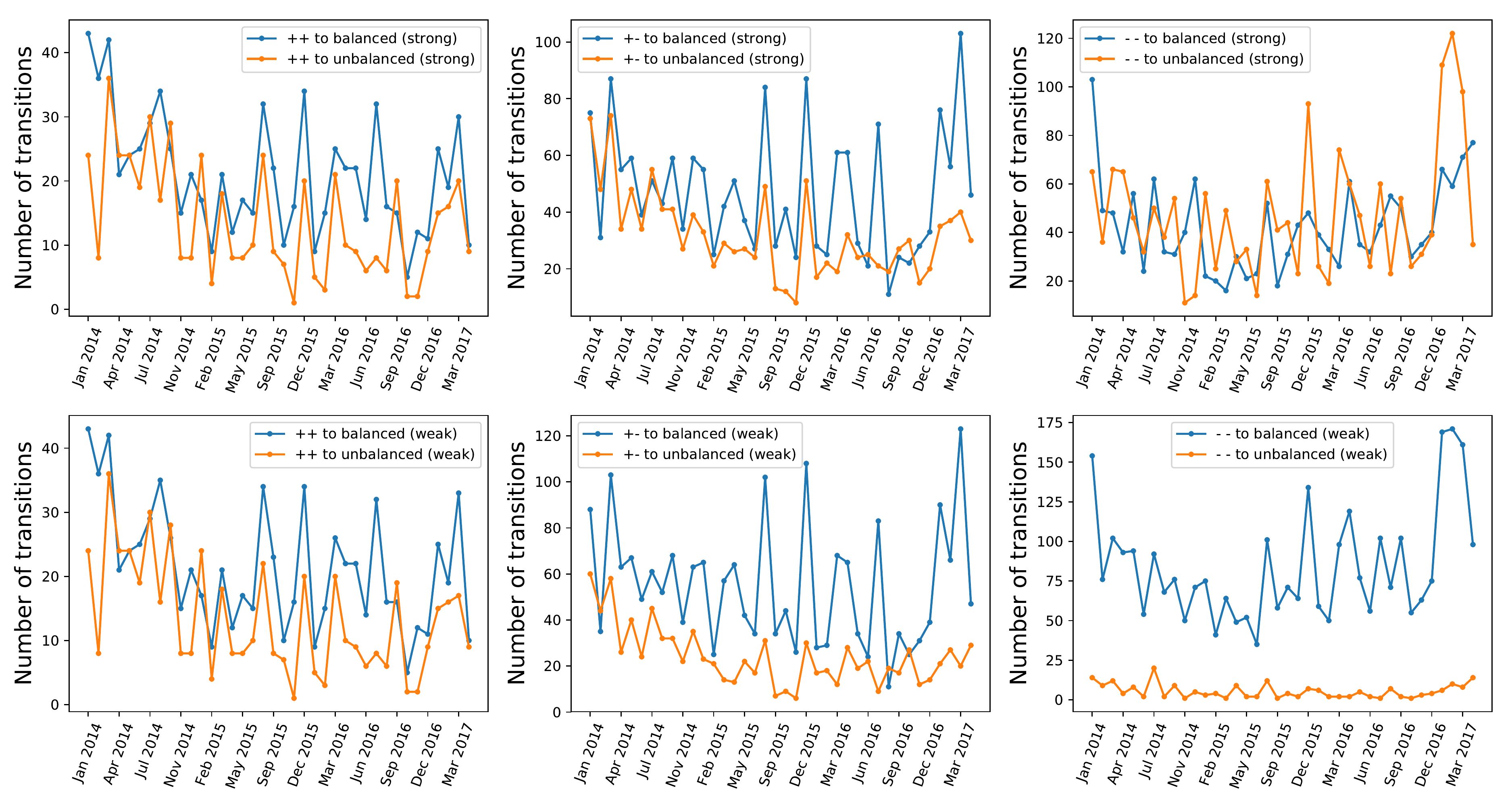}
\caption{
Numbers of transitions from one month to the next, from wedges of each type to balanced and unbalanced triangles, 
where balanced and unbalanced triangles are defined either according to the {strong} (top row) or 
to the weak (bottom row) version of social balance.
}
\label{fig:triadic_closures}
\end{figure}

Moreover, we mention in the main text that the number of triangles and the number of
triadic closure events fluctuate widely, with no clear temporal signal or trend.
This is illustrated both by  Fig. \ref{fig:triadic_closures}
and by  Fig. \ref{fig:avg_clustering_transitivity},  which shows the average
clustering coefficient and the transitivity of each monthly network.

\begin{figure}[h]
\centering
\includegraphics[width=0.5\textwidth]{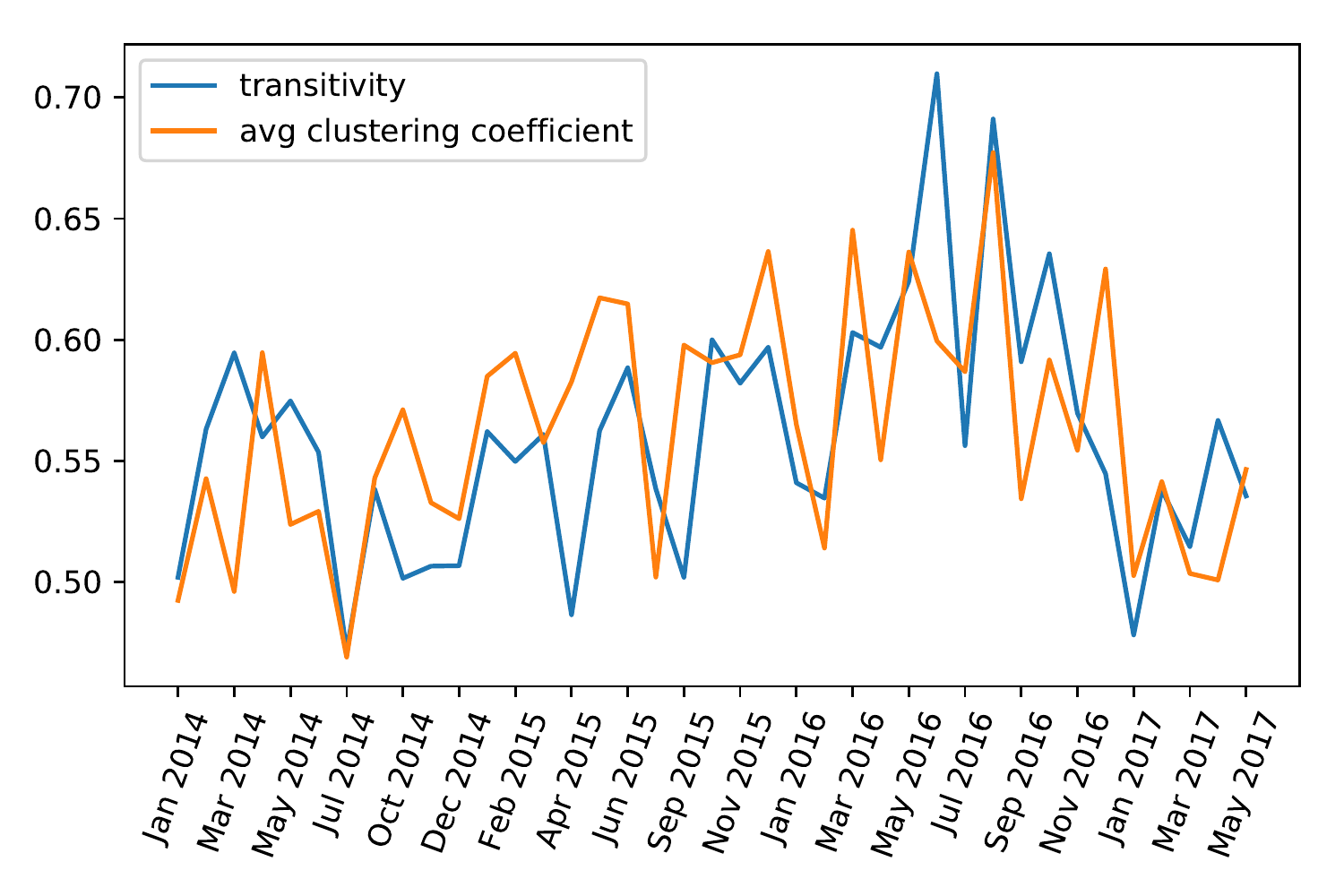}
\caption{\label{fig:avg_clustering_transitivity} 
Transitivity and average clustering coefficient values during the 39 month of investigation.}
\end{figure}

\clearpage
\newpage

\subsection*{Dynamics of individual ego-networks}

Fig. 9 illustrates the evolution of individuals' ego-networks similarities between successive
months. Fig. \ref{fig:ind_trajectories_red_emph_dt_5} displays the same information for all individuals.

\begin{figure}[hb]
\centering
\includegraphics[width=0.85\textwidth]{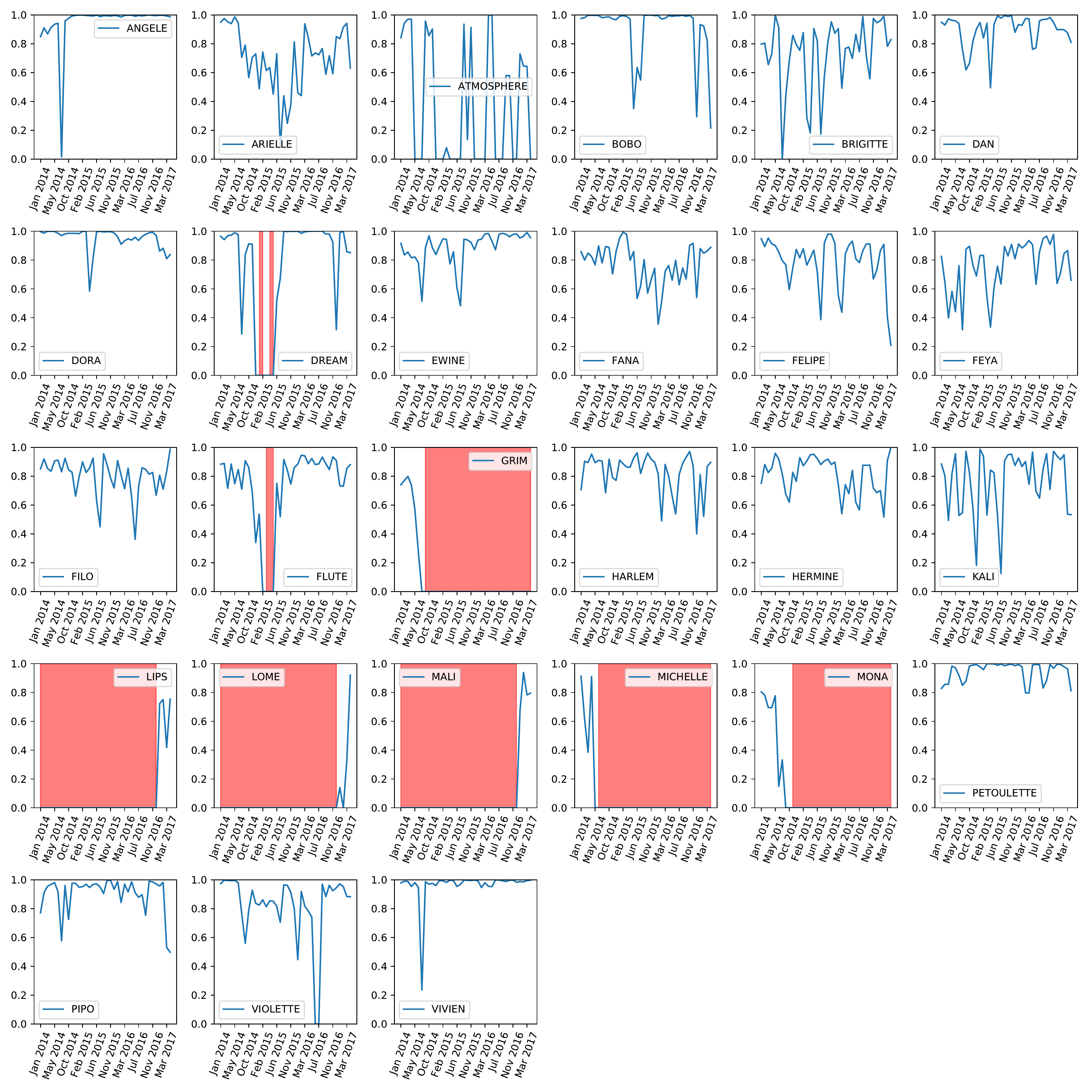}
\caption{\label{fig:ind_trajectories_red_emph_dt_5} 
Evolution of ego-network cosine similarity 
values for all the individuals who were
present at least two successive months during the study.
In each plot, the x-axis corresponds to 
the month
and the y-axis gives the value of the cosine
similarity between the ego-networks of 
the individual in one month and the next. Red vertical bars correspond to the period of absence of that individual from the enclosure.}
\end{figure}

\end{document}